\documentclass[12pt]{iopart}
\usepackage{iopams}  
\usepackage{hyperref}
\usepackage{cite} 
\expandafter\let\csname equation*\endcsname\relax
\expandafter\let\csname endequation*\endcsname\relax

\usepackage{amssymb}
\usepackage{amsmath}
\usepackage{soul}
\usepackage{cite}
\usepackage{enumerate}
\usepackage{empheq}
\usepackage{fancybox}
\usepackage{etoolbox}
\usepackage{hyperref}
\usepackage{float}
\usepackage{color}
\usepackage{xcolor}
\usepackage[normalem]{ulem}
\usepackage{cancel}

\usepackage{orcidlink}

\makeatletter
\def\@mkboth#1#2{}
\newlength\appendixwidth
\preto\appendix{\addtocontents{toc}{\protect\patchl@section}}
\newcommand{\patchl@section}{%
  \settowidth{\appendixwidth}{\textbf{Appendix }}%
  \addtolength{\appendixwidth}{1.5em}%
  \patchcmd{\l@section}{1.5em}{\appendixwidth}{}{\ddt}%
}
\makeatother

\begin{document}

\title[]{Brownian gyrator model with stochastic resetting}

\author{Satya N Majumdar$^1$\orcidlink{0000-0001-9844-6980}, Gr\'egory Schehr$^2\orcidlink{0000-0001-6648-5213}$, Pascal Viot$^3$\orcidlink{0000-0002-1578-1722} \& Gleb Oshanin$^{3}\orcidlink{0000-0001-8467-3226}$}
\address{$^1$ \, Laboratoire de Physique Th\'eorique
et Mod\`{e}les Statistiques (UMR CNRS 8626),
Universit\'e Paris-Saclay,
B\^at. 530, 91405 Orsay Cedex, France}
\address{$^2$ \, Laboratoire de Physique Th\'eorique
	et Hautes Energies (UMR CNRS 7589), Sorbonne Universit\'e, 4 place Jussieu,
	75252 Paris Cedex 05, France}
\address{$^3$ \, Laboratoire de Physique Th\'eorique de la Mati\`{e}re Condens\'ee (UMR CNRS 7600), Sorbonne Universit\'e, 4 place Jussieu,
75252 Paris Cedex 05, France}

\date{\today}

\begin{abstract}
	The Brownian gyrator (BG) model provides a minimal yet experimentally relevant example of a system operating far from thermodynamic equilibrium, sustaining steady currents with nonzero vorticity and exhibiting stochastic rotational motion along elliptical orbits.
	In this paper, we study the response of the BG to a fundamentally non-equilibrium perturbation -- stochastic resetting at a constant rate -- either to a fixed point in the plane or to a randomly selected point on a circle. We determine analytically and numerically the steady-state position probability density function (PDF), together with its marginals and moments, which exhibit a resetting-induced non-monotonic dependence on the resetting rate. We also characterize the associated local probability currents and vorticity fields. Furthermore, we investigate the dependence of the mean angular momentum and mean angular velocity on the resetting rate, and numerically analyze the PDFs of these observables. This analysis fully quantifies the complex behavior arising from the competition between two non-equilibrium mechanisms with distinct spatial symmetries.
		\end{abstract}

\section{Introduction}

The Brownian gyrator (BG) is a paradigmatic model of non-equilibrium statistical mechanics which is often regarded as a minimal microscopic heat engine. It describes a particle confined by a harmonic potential and simultaneously coupled to two thermal reservoirs held at \textit{different} temperatures \cite{Exartier1999}. This temperature anisotropy drives the system out of equilibrium and generates a nonvanishing steady-state probability current, producing a non-zero mean torque that causes the particle to undergo a stochastic rotational motion around the origin \cite{Filliger2007}. 
Owing to its analytical tractability and transparent physical interpretation, the BG has been studied extensively in both theory and experiment. Experimental realizations include equivalent electrical circuits \cite{Ciliberto2013,Ciliberto2013a} and optically trapped colloidal particles subjected to anisotropic thermal environments \cite{Argun2016}. These studies have established the BG as a benchmark experimentally relevant system for exploring fundamental properties of non-equilibrium steady states, including sustained probability currents, finite mean angular momentum and angular velocity \cite{Dotsenko2013,Mancois2018,Bae2021,Chang2021,Siches2022,Dotsenko2024,Mazzolo2023}, and the notion of effective temperatures \cite{Cerasoli2018,Tyagi2020,Cerasoli2021}.
More recent studies have also considered the spinning of the BG around its center of mass \cite{Dotsenko2022a,Dotsenko2023}, the influence of temporally correlated noises \cite{Nascimento2021,Squarcini2022b}, the impact of dichotomous random  driving\cite{Herbeau2026}  as well as analysed the spectral content of the BG trajectories \cite{Ciliberto2013,Ciliberto2013a,Cerasoli2022} and the effects of the inherent anisotropy of the BG (e.g., its biaxiality) \cite{Dutta2026}. In addition, the full probability density functions (PDFs) of angular momentum and angular velocity have been obtained \cite{Viot2024}, revealing the important role of fluctuations and highlighting the limitations of theoretical analyses based solely on mean quantities.

A fundamentally different route to driving a dynamical system out of equilibrium -- distinct from paradigms such as the BG, where non-equilibrium arises from the absence of a unique temperature -- is provided by stochastic resetting. Originally introduced in the context of search processes and first-passage-time optimization \cite{Evans2011,Evans2011a}, resetting has since emerged as a versatile mechanism for inducing and controlling non-equilibrium behavior (for reviews see \cite{Evans2020,pal2022inspection,Gupta2022,Nagar2023,Pal2019landau}). In particular, it has been shown to strongly alter transport and first-passage properties \cite{Evans2020,reuveni2016optimal,pal2017first,kusmierz2014first,majumdar2015dynamical,bressloff2020directed,bressloff2020search}, as well as the thermodynamic characteristics of stochastic systems \cite{fuchs2016stochastic,Mori2023,olsen2024thermodynamic,gupta2022work}, while also reshaping the spectral content and temporal structure of trajectories \cite{Majumdar2018}.

A broad spectrum of stochastic resetting protocols was studied, reflecting the remarkable richness of the resetting paradigm. These protocols include resetting to a fixed position \cite{Evans2011,Evans2011a,Evans2020,Pal2015}, resetting to randomly distributed locations \cite{Evans2011a,besga2020optimal,Sar2023,Olsen2023}, memory-driven strategies in which the system is returned to previously visited states \cite{boyer2014random,Boyer2017,Boyer2024,Boyer2025,mailler2019random}, non instantaneous resetting \cite{evans2019effects,gupta2021stochastic,bodrova2020resetting,radice2021,dubey2026towards,tal2020experimental} and also continuous rescaling protocols that act multiplicatively on the system's state \cite{Biroli2024,olsen2024partial,di2023time}. In addition to theoretical developments,  resetting was also studied experimentally in a variety of settings involving optical traps \cite{tal2020experimental,besga2020optimal,besga2021dynamical,goerlich2025taming,ginot2026experimental}. Despite their conceptual diversity, these mechanisms share the ability to generate highly nontrivial non-equilibrium steady states that differ fundamentally from their equilibrium counterparts. Such stationary distributions frequently display singular features, such as cusp-like non-analyticities in the probability density near the resetting positions \cite{Evans2011,Evans2011a,evans2014diffusion,Evans2020}.
These stationary states are in general of non-equilibrium nature, carrying a nonzero probability current in the configuration space, originating from the violation of detailed balance by the resetting moves \cite{Evans2011,Evans2011a,Evans2020}.

In this paper, we bring together these two active areas of research, namely the Brownian gyrator and the stochastic resetting, by studying the BG model subjected to stochastic resetting. Specifically, we consider Poissonian resetting either to a fixed point in the plane or to a point selected uniformly at random on a circle centered at the origin. These two protocols allow us to examine how the rate and spatial geometry of resetting affect the stationary state and, in particular, the BG rotational dynamics.

Our motivation is twofold. First, as demonstrated in \cite{Viot2024}, the PDF of the angular velocity in the conventional BG exhibits algebraically decaying tails, such that only its first moment is finite. Such anomalously large fluctuations lead to highly intermittent rotational dynamics and raise questions about the robustness of the BG as a microscopic engine. It is therefore natural to ask whether stochastic resetting
 can regularize these fluctuations by interrupting trajectories and limiting their excursions. We show that it cannot: the algebraic tails of the PDF of angular velocity persist in the presence of resetting. This robustness highlights the distinction between the fluctuations of angular velocity, which are particularly sensitive to passages close to the origin, and those of angular momentum, whose large values are associated with large spatial excursions.

Second, thermal non-equilibrium driving of the standard BG model and stochastic resetting are two distinct mechanisms for maintaining a system out of equilibrium, with different spatial symmetries. Their interplay raises the broader question of whether resetting merely changes the strength of the gyration or can qualitatively reorganize its rotational motion. We find that the latter is indeed possible. For resetting to a fixed point, the competition between thermal anisotropy and the spatial asymmetry of the resetting position can even reverse the sign of the mean angular momentum. Moreover, resetting generates nontrivial local probability currents and spatially structured vorticity fields, even when the two temperatures are equal and the underlying BG dynamics would satisfy detailed balance. These local flow patterns therefore reveal a richer organization of rotational motion than can be captured by global mean observables alone.

The resetting geometry also plays a decisive role. While resetting to a fixed point introduces a preferred spatial direction, uniform resetting on a circle restores the symmetry of the resetting protocol under rotations about the origin. As a consequence, the two protocols affect rotational observables in markedly different ways: in particular, circular resetting suppresses the mean angular momentum without inducing the reversal possible for fixed-point resetting, whereas the mean angular velocity retains a nontrivial dependence on the resetting radius. In addition, resetting can induce correlations between otherwise dynamically decoupled coordinates and produce non-monotonic dependencies of correlations and position fluctuations on the resetting rate. Together with the emergence of non-Gaussian and non-elliptic stationary position distributions, these effects demonstrate that resetting provides a means of reshaping both the spatial structure and the rotational dynamics of the non-equilibrium steady state.

	The remainder of the paper is organized as follows. In Sec.~\ref{sec:BG}, we introduce the standard BG model and establish the notation used throughout the paper. Section~\ref{sec:resetBG} is devoted to the analysis of Poissonian resetting to a fixed position, while Sec.~\ref{sec:resetBG2} examines resetting to random points uniformly distributed on a circle. Finally, Sec.~\ref{sec:conclusions} 
	summarizes our main findings and discusses possible directions for future research. The details of  intermediate calculations  and some of our results are relegated to Appendices.

\section{The model}\label{sec:BG}
The standard BG model (in the absence of resetting) consists of two linearly coupled stochastic differential equations obeyed by two functions, $x_t$ and $y_t$ \cite{Exartier1999,Filliger2007}:
\begin{align}
\begin{split}
\label{Lan}
 \dot{x}_t &= - x_t - u y_t + \xi_x(t) \,,  \\
 \dot{y}_t &= - y_t - u x_t + \xi_y(t) \,,  
\end{split}
\end{align}
where the dot denotes the time-derivative and $\xi_x$ and $\xi_y$ are Gaussian noises with zero mean and the covariance functions
\begin{align}
\begin{split}
\label{noises}
& \overline{\xi_x(t) \xi_x(t')} = 2 T_x \delta(t-t') \,, \\
& \overline{\xi_y(t) \xi_y(t')} = 2 T_y \delta(t-t') \,, \\
& \overline{\xi_x(t) \xi_y(t')} = 0 \;.
\end{split}
\end{align}
The overbar here and henceforth denotes averaging over thermal histories. The noise terms acting on 
$x_t$ and 
$y_t$ are statistically independent and, in general, their amplitudes (temperatures $T_x$ and $T_y$) are not equal to each other.  
Since $\xi_x(t)$ and $\xi_y(t)$ are Gaussian noises and the equations of motion \eqref{Lan} are linear, it follows that $x_t$ and $y_t$ will be jointly Gaussian. We will derive this bivariate Gaussian distribution explicitly at all times $t$.

The parameter $u$ in Eqs. \eqref{Lan} denotes the coupling between $x_t$ and $y_t$. 
 Importantly, a stationary solution exists only if $|u| < 1$. Once $|u| \geq 1$, the dynamics is no longer confined and the system does not approach a steady state. In the particular case $u=0$, the dynamics reduces trivially to two independent Ornstein–Uhlenbeck processes each evolving  at  its own temperature. For $u \neq 0$  the two processes  are coupled and once $T_x \neq T_y$ the system converges towards a non-equilibrium steady-state.

The functions $x_t$ and $y_t$ can also be identified as the instantaneous position coordinates of a particle moving on the $(x,y)$-plane, in a parabolic potential
\begin{equation}
U = \frac{1}{2} \left(x^2 + y^2\right) + u x y \,, 
\end{equation} 
subject to noises with temperatures $T_x$ and $T_y$ acting along the $x$ and $y$-axes, respectively.

For the standard BG model as defined above, the PDF $P(x,y;t)$ for finding the particle at position $(x,y)$ at time instant $t$ obeys
the Fokker-Planck equation of the form
\begin{equation}
\label{1}
\dot{P}(x,y;t) = \hat{{\cal L}} P(x,y;t) \,,
\end{equation}
where the operator $\hat{{\cal L}}$ is given by
\begin{equation}
\hat{{\cal L}} = \frac{\partial }{\partial x} \left( T_x \frac{\partial}{\partial x}  + \left(x + u y \right) \right) + \frac{\partial }{\partial y} \left( T_y \frac{\partial}{\partial y}  + \left(y + u x \right) \right)  \,.
\end{equation}
In the steady-state, the solution of Eqs.  \eqref{1} reads
\begin{equation}
\label{P}
P^{\rm st}(x,y) = Z^{-1} \exp\left(- A \left(\left(T_y + \frac{u^2 \Delta_T}{2}\right) x^2
+ \left(T_x - \frac{u^2 \Delta_T}{2}\right) y^2
+ u \left(T_x + T_y\right) x y
\right)\right) \,,
\end{equation}
where the superscript "st" here and henceforth indicates that we consider the position PDF in the limit $t = \infty$; $A$, $\Delta T$ and the normalization $Z^{-1}$  are explicitly given by
\begin{align}
& A = \frac{2}{4 T_x T_y + u^2 \Delta_T^2}  \,, \quad \Delta_T = T_x - T_y \,, \quad
Z= \pi \sqrt{\frac{4 T_x T_y + u^2 \Delta_T^2}{1 - u^2}} \,.
\end{align} 
One notices that for $T_x \neq T_y$ the PDF in Eq. \eqref{P} has elliptic symmetry, with elliptical iso-probability contours.  Such elliptic contours are apparent in Fig. \ref{fig:1}(a) below. We also note that, as discussed above, the normalization $Z$ is finite only for $|u| < 1$.

The aim of the present paper is to generalize the standard BG model by introducing stochastic Poissonian resetting with a constant rate 
$r$ \cite{Evans2011,Evans2011a}. Within this framework, the dynamics of the BG is intermittently interrupted by resetting events occurring at random times, at which the BG is instantaneously returned to a prescribed spatial location. 
The time-intervals between reset events are independent, exponentially distributed random variables. 
We consider two distinct resetting protocols. In the first scenario (case $A$), the BG is reset to a fixed point with (arbitrary) coordinates $(X_0,Y_0)$. In the second scenario (case $B$), the BG is reset to a position independently and uniformly sampled along a circle of fixed radius $R_0$ centered at the origin, with probability density $1/(2\pi R_0)$. This resetting protocol introduces an additional circular symmetry into the model.

\section{Case $A$: Resetting to a fixed point in the plane}\label{sec:resetBG}

In this section, which introduces basic concepts and notations, we discuss the dynamics of the BG in the presence of stochastic resetting with a constant rate 
$r$ to a fixed point 
$(X_0,Y_0)$ in the 
$(x,y)$-plane. For clarity of exposition, we outline this section as follows: We begin by recalling the fundamental renewal relation that connects the evolution of the system in the absence of resetting with that in the presence of stochastic resetting events. This representation allows us to define the stationary position PDF of the BG under resetting and to analyse its asymptotic behavior. On this basis, we subsequently compute the moments of this PDF and show that they exhibit nontrivial non-monotonic dependencies on the resetting rate. We then present a numerical analysis of the mean angular momentum and the mean angular velocity of the BG. Finally, we investigate the structure of the local probability currents and  their vorticity fields.

\subsection{Renewal representation for stochastic resetting}

Consider a stochastic process $\mathbf{X}(t)$ evolving in  $d$-dimensional space in continuous time.  Assume that this process is characterized, in the absence of resetting, by the position PDF $P(\mathbf{X};t|\mathbf{X}_0)$, i.e., the probability density of finding $\mathbf{X}(t)$ at position $\mathbf{X}$ at time $t$, given that it started from $\mathbf{X}_0$ at time $t=0$. We perturb then the process by stochastic resetting to the fixed position $\mathbf{X}_0$. Resetting events occur at random times generated by a Poisson process with constant rate $r$, so that the waiting times between successive resetting events are independent  exponentially distributed random variables with density $r e^{-r\tau}$. Let $P_r(\mathbf{X};t)$ now denote the probability density of finding $\mathbf{X}(t)$ at position $\mathbf{X}$ at time $t$ in the presence of resetting (indicated by the subscript "r") to a fixed point $\mathbf{X}_0$.

Because the resetting protocol is memoryless, the process admits a renewal representation \cite{Evans2011,Evans2011a,Bray2013} in which the PDF at time $t$ is decomposed according to the time elapsed since the most recent resetting event. In particular, two classes of trajectories contribute to $P_r(\mathbf{X},t)$: those for which no resetting has occurred up to time $t$, and those for which the last resetting event occurred at time $t-\tau$, followed by a period of free evolution of duration $\tau$. Since the probability that no resetting occurs up to time $t$ is $e^{-rt}$, this decomposition yields the following  renewal equation \cite{Evans2011,Evans2011a,Bray2013}
\begin{align}
	P_r(\mathbf{X};t)
	=
	e^{-rt} P(\mathbf{X};t|\mathbf{X}_0)
	+
	r \int_0^t d\tau \, e^{-r\tau} P(\mathbf{X};\tau|\mathbf{X}_0) \,.
\end{align}
This fairly general relation expresses the propagator of the resetting process in terms of the propagator of the underlying stochastic dynamics without resetting.  The renewal structure embodied in this equation constitutes the starting point for the analysis of many statistical properties of stochastic processes with resetting, including stationary states, relaxation dynamics, and first-passage observables.

In the long-time limit, provided the integral converges, the process reaches a non-equilibrium stationary state. Taking the limit $t \to \infty$ in the renewal equation and noting that the contribution from trajectories without resetting, proportional to $e^{-rt}$, vanishes exponentially, we obtain
\begin{equation}
	\label{9}
	P_r^{\rm st}(\mathbf{X})
	=
	r \int_0^\infty d\tau \, e^{-r\tau} P(\mathbf{X};\tau|\mathbf{X}_0) \,.
\end{equation}
This expression shows that the stationary distribution in the presence of resetting can be interpreted as a weighted average of the propagator of the underlying stochastic process, where the weights correspond to the probability density of the time elapsed since the most recent resetting event. The exponential kernel $r e^{-r\tau}$ reflects the Poissonian statistics of the resetting events and emphasises shorter excursions away from $\mathbf{X}_0$.

\subsection{Position PDF of the BG with resetting to a fixed point}

The position PDF of the BG starting at $(X_0,Y_0)$  at $t = 0$ and evolving in the absence of stochastic resetting can be formally represented as
\begin{align}
	\label{psir}
	P(x,y;t|X_0,Y_0) = \overline{\delta(x - x_t) \delta(y - y_t)} \,, \quad  (x_0,y_0) = (X_0,Y_0) \,,
\end{align}
where  $x_t$ and $y_t$ are the solutions of the Eqs. \eqref{Lan}
for  given realizations of noises, 
\begin{align}
	\begin{split}
		x_t &= X_0 \, e^{-t} \cosh(u t) - Y_0 \, e^{-t} \sinh(u t) \\
		&+ e^{- t} \int^t_0 d\tau \, e^{\tau} \, \cosh\left(u (t - \tau)\right) \, \xi_x - e^{- t} \int^t_0 d\tau \, e^{\tau} \, \sinh\left(u (t - \tau)\right) \, \xi_y \,, \\
		y_t &= Y_0 \, e^{-t} \cosh(u t) - X_0 \, e^{-t} \sinh(u t) \\
		&- e^{- t} \int^t_0 d\tau \, e^{\tau} \, \sinh\left(u (t - \tau)\right) \, \xi_x + e^{- t} \int^t_0 d\tau \, e^{\tau} \, \cosh\left(u (t - \tau)\right) \, \xi_y \,. 
	\end{split}
\end{align}
Averaging in Eq. \eqref{psir} is performed straightforwardly to give
\begin{align}
	\begin{split}
		\label{l}
	&	P(x,y;t|X_0,Y_0) =  \frac{1}{(2 \pi)^2} \int^{\infty}_{-\infty} \int^{\infty}_{-\infty} d\omega_1 \, d\omega_2 \, \overline{\exp\left(i \omega_1 (x - x_t) +  i \omega_2 (y - y_t)\right)} \\
		&= \frac{1}{2 \pi \sqrt{d}} \exp\left(- \Psi\right) \,, \quad \Psi = \frac{b \, (x - \tilde{X}_0)^2 + a \, (y - \tilde{Y}_0)^2 + c \, (x - \tilde{X}_0) (y - \tilde{Y}_0)}{d} \,,
	\end{split}
\end{align}
where $d = 4 a b - c^2$ and $\tilde{X}_0$, $\tilde{Y}_0$,  as well as $a$, $b$ and $c$ are the following functions of time:
\begin{align}
	\begin{split}
		\label{tilde}
		\tilde{X}_0 &= X_0 \, e^{-t} \cosh(u t) - Y_0 \, e^{-t} \sinh(u t) \,, \quad 
		\tilde{Y}_0  = Y_0 \, e^{-t} \cosh(u t) - X_0 \, e^{-t} \sinh(u t) \,,
	\end{split}
\end{align}
and 
\begin{align}
	\begin{split}
		\label{coeft}
		a& =  a_t = \frac{1}{4 (1 - u^2)} \Big( 2 T_x - u^2 \Delta_T \, - \\ &- (T_x+ T_y) \, \left(\cosh(2 u t) + u \sinh(2 u t)\right) \, e^{-2 t} -  (1 - u^2) \Delta_T \, e^{-2 t} \Big) \,, \\ 
		b& = b_t = \frac{1}{4 (1 - u^2)} \Big( 2 T_y + u^2 \Delta_T \, - \\ &- (T_x+ T_y) \, \left(\cosh(2 u t) + u \sinh(2 u t)\right) \, e^{-2 t} + (1 - u^2) \Delta_T \, e^{-2 t} \Big) \,, \\ 
		c&= c_t =  \frac{(T_x + T_y)}{2 (1 - u^2)} \Big(u - \left(u \cosh(2 u t) + \sinh(2 u t)\right) e^{- 2 t} \Big) \,. 
	\end{split}
\end{align}
For brevity, we henceforth omit the subscript $t$ when referring to these functions.

Consequently, the position PDF  
$P(x,y;t|X_0,Y_0)$  of the BG in the absence of resetting is fully determined for all times 
$t$. Note that Eqs. \eqref{l}-\eqref{coeft} are valid at all times $t$ and for arbitrary $u$. However, if one takes the limit $t \to \infty$, keeping $u$ fixed, the stationary state exists only for $|u|<1$. Note that, for finite $t$, the functions $a, b$ and $c$ in Eqs. \eqref{coeft} may naively seem divergent, but one can check that in the limit $u \to 1$ they remain finite as long as $t$ is finite.

Consider now the situation where a finite resetting rate $r$ is switched on. According to Eq. \eqref{9}, to compute the stationary joint distribution in the presence of resetting, we need the full time-dependent probability distribution in the absence of resetting. The latter is derived in Eqs. \eqref{l}-\eqref{coeft}. Substituting this result in Eq. \eqref{9} gives us the stationary joint PDF in the presence of resetting. We get
\begin{align}
	\label{pr2}
	P^{\rm  st}_r(x,y|X_0,Y_0) = \frac{r}{2 \pi } \int^{\infty}_0 \frac{e^{- r \tau} \, d\tau}{\sqrt{d}} \exp\left(- \Psi\right) \,, 
\end{align} 
where $\Psi$ is defined  in Eq. \eqref{l}. Note that, while the reset free distribution at any finite time $t$ is Gaussian where $u$ is arbitrary, the distribution in the presence of resetting in Eq. \eqref{pr2} is not Gaussian and is valid for all $u$. Due to a complicated time-dependence of the functions $a$, $b$ and $c$, the Laplace transform in Eq. \eqref{pr2} cannot be performed in an explicit form. We will therefore try first to get an idea about the shapes of $P^{\rm st}_r(x,y|X_0,Y_0)$ by performing the integral in Eq. \eqref{pr2} numerically, and then derive explicit expressions for the asymptotic behavior of $P^{\rm st}_r(x,y|X_0,Y_0)$ in the limit  $r \to \infty$.

\begin{figure}[h]
	\centering
	\includegraphics[width=51mm]{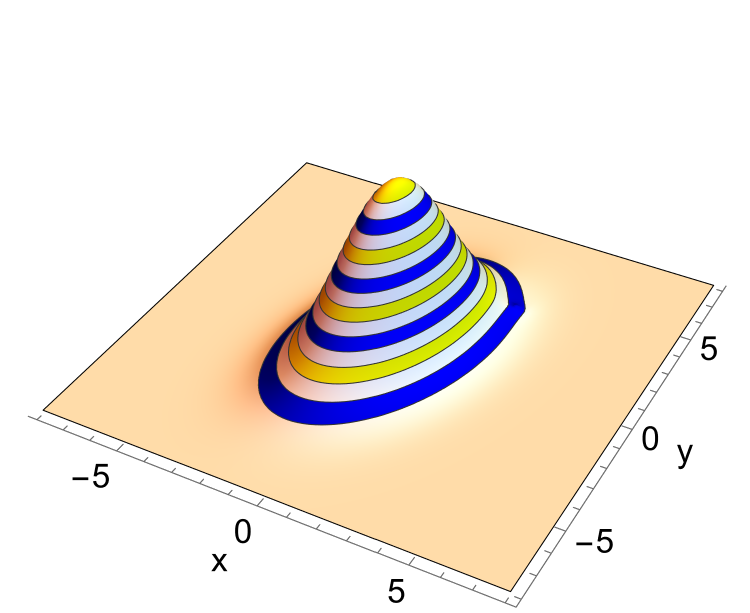}  
	\includegraphics[width=51mm]{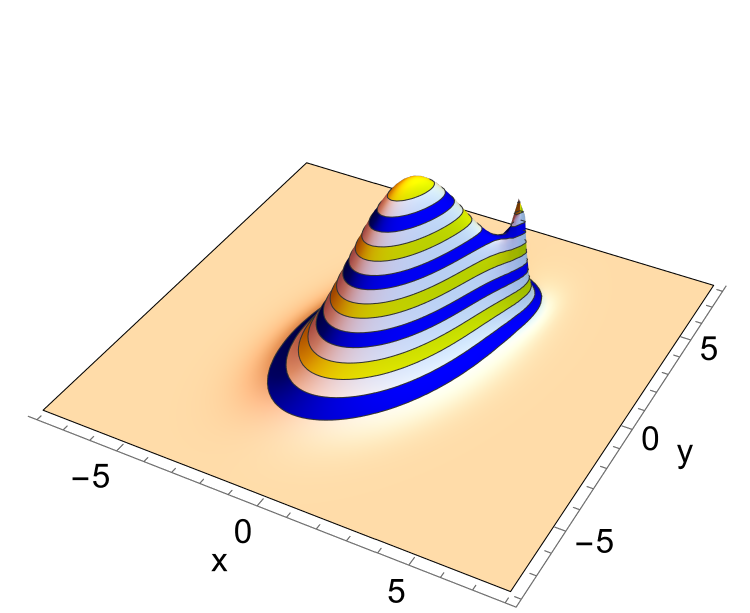}
	\includegraphics[width=51mm]{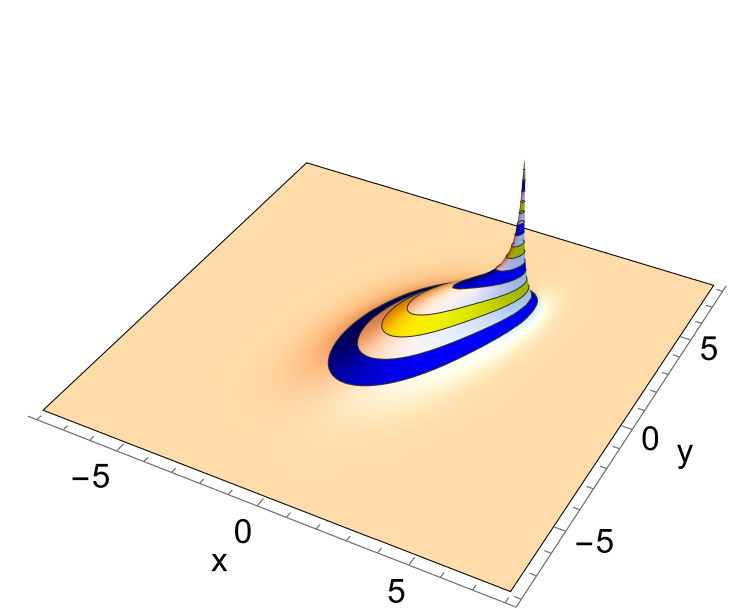}
	\caption{Case A. The probability density function $P^{\rm  st}_r(x,y|X_0,Y_0)$ in Eq. \eqref{pr2} versus $x$ and $y$. The resetting point is at $X_0 = Y_0 = 3$, the coupling parameter $u = 0.5$, and the temperatures $T_x = 1$ and $T_y = 4$. Left panel: $r = 0$ (see Eq. \eqref{P}, no resetting). Middle panel: $r = 0.2$. Right panel: $r = 0.7$. As the resetting rate 
		$r$
	increases, the PDF progressively loses its elliptic symmetry and acquires a bimodal form.}
	\label{fig:1} 
\end{figure}

Figure ~\ref{fig:1}   presents density plots of 
$P^{\rm  st}_r(x,y|X_0,Y_0)$ 
as functions of 
$x$
and 
$y$, obtained from a numerical evaluation of the integral in Eq. \eqref{pr2} for unequal temperatures 
$T_x=1$ and
$T_y=4$, and for three values of the resetting rate: 
$r=0$, 
$r=0.2$ 
and 
$r=0.7$. The left panel shows  the “pure” case 
$r=0$, corresponding to the standard BG model. The elliptic symmetry of the distribution is clearly visible. The middle panel displays the case of weak resetting, 
$r=0.2$. Here, resetting acts as a small perturbation, giving rise to a secondary peak at the resetting location (here $X_0=Y_0=3$), whose height remains lower than the main peak at the origin characteristic of the BG dynamics. In the right panel, where the resetting rate is increased to 
$r=0.7$, the peak at the resetting point becomes dominant, while the central structure associated with the standard BG model is suppressed. This demonstrates that already for such moderate values of 
$r$ resetting can no longer be regarded as a minor perturbation: it induces a substantial qualitative change in the stationary distribution and strongly alters the gyration process. As we show below, this effect is also reflected in the behavior of dynamical observables.

Large-$r$ asymptotic behavior of $P^{\rm  st}_r(x,y|X_0,Y_0)$ can be accessed as follows. 
For large values of $r$, the exponential function  $\exp(- r \tau)$ suppresses completely the large-$\tau$ behavior of the integrand. 
 Therefore, it suffices to retain only the leading-order terms in the short-$\tau$ expansions of the functions $a, b, c$ and $d$ defined in Eqs. \eqref{coeft},
\begin{align}
	\begin{split}
		\label{lowt1}
a &= T_x \tau -T_x \tau^2 +O(\tau^3),\quad 
b = T_y \tau -T_y \tau^2+O(\tau^3),\quad \\
c &= u (T_x + T_y)\tau^2 + O(\tau^3) ,\quad d= 4 T_x\, T_y \,\tau^2 + O(\tau^3)
	\end{split}
\end{align}
while the  functions $\tilde{X}_0$ and $\tilde{Y}_0$ in this limit follow
\begin{align}
		\label{lowt2}
		\tilde{X}_0 &= X_0 - (X_0 + u Y_0)  \tau + O(\tau^2) \,,\quad
		\tilde{Y}_0 = Y_0 - (u X_0 + Y_0)  \tau + O(\tau^2)  \,.
\end{align}
Inserting the above asymptotic forms  into  Eq. \eqref{pr2} and integrating over $\tau$, we find that in the large-$r$ limit the position PDF attains the following asymptotic form
\begin{align}
	\begin{split}
		\label{larger}
		P^{\rm  st}_r(x,y&|X_0,Y_0) \simeq \frac{r}{2 \pi \sqrt{T_x T_y}} \exp\left(- \frac{(X_0 + u Y_0)(x - X_0)}{2 T_x} - \frac{(u X_0 + Y_0)(y - Y_0)}{2 T_y}\right)  \\
		& K_0\left( \sqrt{\left(r + \frac{(u X_0 + Y_0)^2}{4 T_y} + \frac{(X_0 + u Y_0)^2}{4 T_x}\right) \left(\frac{(x - X_0)^2}{T_x} + \frac{(y - Y_0)^2}{T_y} \right)}  \right) \,,
	\end{split}
\end{align}
where $K_0(z)$ is the modified Bessel function of the second kind \cite{AbramowitzStegun1964} 
Consequently, 
in the limit of a very frequent resetting  the PDF $P^{\rm  st}_r(x,y|X_0,Y_0)$ is no longer a Gaussian function of $x$ and $y$, and vanishes with $r$ exponentially, 
as $\exp(-\sqrt{r})$, for $x$ and $y$ bounded away from $X_0$ and $Y_0$. In turn,
we observe that  $P^{\rm  st}_r(x,y|X_0,Y_0)$ in Eq. \eqref{larger} is \textit{logarithmically} diverging at the resetting point, i.e., when $x=X_0$ and $y = Y_0$. This is reminiscent of the two-dimensional diffusion in the presence of stochastic resetting~\cite{evans2014diffusion}. Note that, for large $r$ and $X_0=Y_0=0$ (frequent reset to the origin) and for simplicity setting $T_x = T_y = T$, the result in Eq. \eqref{larger} reduces to
\begin{align} \label{reset_origin}
P^{\rm  st}_r(x,y|X_0 = 0,Y_0 = 0) \simeq \frac{r}{2 \pi T} K_0 \left( \sqrt{\frac{r}{T} (x^2+y^2) }\right) \;,
\end{align}
which coincides with the result for two-dimensional diffusion with stochastic resetting for arbitrary $r$. This result is completely independent of the coupling parameter $u$. This can be understood as follows. In the limit of large $r$ and resetting to the origin, the particle mostly diffuses very close to the origin and does not feel the external deterministic force due to the confinement and the coupling. Thus the large $r$ limit in the presence of confinement and coupling becomes equivalent to the standard two-dimensional diffusion with stochastic resetting in Eq. \eqref{reset_origin} with rate $r$.
\begin{figure}[h]
	\begin{center}
		\includegraphics[width=80mm]{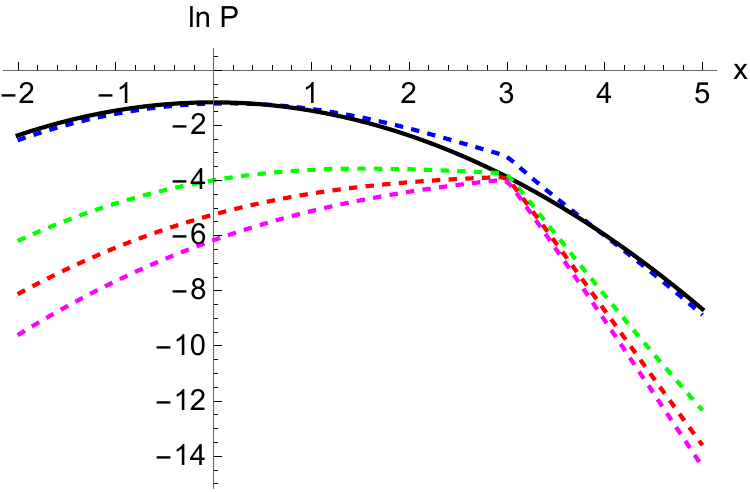}  
	\end{center}
	\caption{Case A. Numerically evaluated marginal probability density function $P^{\rm st}_r(x|X_0,Y_0)$ in Eq. \eqref{p} as a function of $x$ for $X_0=Y_0 = 3$, $T_x = 1$, $T_y = 3$ and $u = 1/2$. The black solid curve shows the result in the absence of resetting, i.e., for 
		$r=0$. The dashed blue, green, red, and magenta curves correspond to 
		$r=0.1$,
		$r=1$, $r=2$ and 
		$r=3$, respectively. As the resetting rate 
		$r$
		increases, the marginal PDF progressively develops a cusp at 
		$x=X_0$ (here 
		$X_0=3$).}
	\label{fig:X} 
\end{figure}
The \textit{marginal} distribution along one of the axes, say, along the $X$-axis, is given by
\begin{align}
	\begin{split}
	\label{p}
	P^{\rm st}_r(x|X_0,Y_0) &= \int^{\infty}_{-\infty} dy \, P^{\rm st}_r(x,y|X_0,Y_0) \\&= \frac{r}{2 \sqrt{\pi}} \int^{\infty}_0 \frac{d\tau}{\sqrt{a}} \exp\left(- r \tau - \left(x - \tilde{X}_0\right)^2/4 a \right) \,,
	\end{split}
\end{align}
where $\tilde{X}_0$ and $a$ are defined in Eqs.  \eqref{tilde} and \eqref{coeft}. Using the asymptotic short-$\tau$ expansions of $\tilde{X}_0$ and $a$ presented in Eqs.  \eqref{lowt1} and \eqref{lowt2}, we find that 
in the large-$r$ limit the marginal PDF attains the following exponential form 
\begin{align}
	\begin{split}
		\label{pp}
	&P^{\rm st}_r(x|X_0,Y_0) \simeq \frac{r}{\sqrt{4 T_x r + (X_0 + u Y_0)^2}} \\
	&\times  \exp\left(- \frac{(X_0 + u Y_0)}{2 T_x} (x - X_0) - \frac{\sqrt{4 T_x r + (X_0 + u Y_0)^2}}{2 T_x} |x - X_0|\right) \,.
	\end{split}
\end{align}
The above function has a cusp at $x = X_0$. Note that such a cusp is already apparent in Fig.~\ref{fig:X} for rather small values of the resetting rate 
$r$, suggesting that it is a generic feature of the  marginal PDF. We also emphasise that the marginal PDF in Eqs. ~\eqref{p} and \eqref{pp} depends parametrically on the 
$y$-coordinate of the resetting point through the dependence of 
$\tilde{X}_0$
on $Y_0$. Note that $P_r^{\rm st}(x|X_0,Y_0)$ drops with $r$ in proportion to $\exp(- \sqrt{r})$, exactly as the full PDF in Eq. \eqref{larger}.

\subsection{Moments and cross-moments of the position PDF}

To derive explicit expressions for the moments and cross-moments of the position PDF in the presence of resetting to a fixed point, it is convenient to consider instead the characteristic function, which satisfies 
\begin{align}
	\begin{split}
		\label{Phi}
		\Phi_r^{\rm st}(\omega_1,\omega_2&|X_0,Y_0) =  \int^{\infty}_{-\infty} \int^{\infty}_{-\infty} dx \, dy \, \exp\left(- i \omega_1 x - i \omega_2 y\right) P_r^{\rm st}(x,y|X_0,Y_0) \\
		& = r \int^{\infty}_0 d\tau \, e^{- r \tau} \, \exp\left(- i \omega_1 \tilde{X}_0 - i \omega_2 \tilde{Y}_0 - a \, \omega_1^2 + c  \, \omega_1 \omega_2 - b \, \omega_2^2  \right) \,.
	\end{split}
\end{align}
The moments and cross-moments can then be obtained by appropriate differentiation of this expression, followed by setting $\omega_1 = \omega_2 = 0$. In doing so, 
we find rather simple expressions for 
	the first moments of positions
	\begin{align}\label{eq:mom}
		\begin{split}
			\overline{x} = \frac{r \left(X_0 + r X_0 - u Y_0\right)}{(1+r)^2 - u^2} \,, \quad
			\overline{y} = \frac{r \left(Y_0 + r Y_0 - u X_0\right)}{(1+r)^2 - u^2} \,.
		\end{split}
	\end{align}  
while the lowest-order cross-moment obeys
		\begin{align}\label{eq:mom1}
		\overline{x y} = \frac{r (2+r) X_0 Y_0 - u r (X_0^2 + Y_0^2) - 2 u (T_x + T_y)}{(2 + r)^2  - 4 u^2} \,,
	\end{align}
	and shows quite a non-trivial dependence 
	on the parameters of the model.

It is instructive to emphasize the following
three points:
	\begin{itemize}
	\item The first moments of positions vanish when $r = 0$ because the system reverts to a purely elliptically symmetric state, whereas  $ \overline{x}\to X_0$ and $ \overline{y}  \to Y_0$ in the opposite limit $r \to \infty$, as they should. 
	The first cross-moment does not vanish when $r = 0$, because in the standard BG model the $x$- and $y$-components of its position become correlated due to the coupling parameter $u$.  
	\item 
	The first moments and the cross-moment (as well as higher-order moments, see below) remain finite for 
	$|u|<1+r$. This condition is less restrictive than in the standard BG model without resetting, for which the existence of finite moments requires 	$|u|<1$. Hence, stochastic resetting effectively enlarges the parameter range over which a well-defined steady state exists, extending it to values of 
	$u$ satisfying 	$|u| < 1+r$.
	This observation highlights a key effect of resetting: it stabilises the dynamics by counteracting the tendency toward delocalisation, thereby significantly modifying the properties of the non-equilibrium steady state.
	\item Focus next on the dependence of the correlation function
	\begin{align}
	\label{corr2}
		{\rm corr}(x,y) = \overline{x y} - \overline{x } \, \overline{y}
			\end{align}
	on the resetting rate $r$. This function equals $- u (T_x +  T_y)/2  (1-u^2)$ for  $r = 0$,  and  vanishes when  $r  \to \infty$. We start with the simplest case when $u = 0$, i.e., when the $x$- and $y$-components of the BG in Eqs. \eqref{Lan} are \textit{decoupled}  in the absence of resetting. In this case we get  from Eqs. \eqref{eq:mom} and  \eqref{eq:mom1}
	\begin{align}
	\label{corr}
	{\rm corr}(x,y) = \frac{r X_0 Y_0}{(2+r)(1+r)^2} \,.
	\end{align} 
	A remarkable feature of this simple result is that 
	${\rm corr}(x,y) \neq  0$  for finite $r$, indicating that stochastic resetting induces correlations between the two components even though they are otherwise dynamically decoupled. In the absence of resetting ($r = 0$), the 
	$x$- and 
	$y$-coordinates evolve independently when $u=0$. However, resetting introduces a common random sequence of events that simultaneously affects both degrees of freedom. In other words, the resetting events act as a shared stochastic “clock”, imposing a global conditioning on the dynamics. As a consequence, variables that would otherwise remain statistically independent become correlated in the stationary state.
	Therefore, the model defined in Eqs. ~\eqref{Lan}, even in the case $u=0$,
	provides a particularly transparent example of a system in which stochastic resetting generates correlations purely through the presence of \text{common} resetting events. The resulting stationary state is thus correlated despite the absence of direct dynamical coupling. This mechanism,  called the \emph{dynamically emergent correlations} (DEC) has attracted considerable attention in a number of recent studies, both theoretically \cite{biroli2023extreme,sabhapandit2024noninteracting,Biroli2024a,Kulkarni2025,mesquita2025dynamically,majumdar2026dynamically,Mauro2026,de_Mauro_2026} and experimentally~\cite{biroli2026experimental}. 
	
	Another salient feature of the result in Eq. ~\eqref{corr} is the \textit{non-monotonic} dependence of the correlations on the resetting rate. In particular, one has 
	${\rm corr}(x,y)=0$
	both in the absence of resetting ($r=0$) and in the limit of infinitely frequent resetting ($r \to \infty$). This implies the existence of an intermediate value 
	$r=r^* = (\sqrt{5}-1)/2$
	at which the correlation attains a maximal magnitude. Physically, this behavior reflects a competition between two effects: for small $r$, resetting events are too rare to induce significant correlations, whereas for very large $r$, the dynamics is dominated by frequent returns to the resetting point, which suppress  correlations.
				
	As we briefly demonstrate below, this non-monotonic dependence is not restricted to the case 
			$u=0$.
		 It can also arise for 
			$u \neq 0$;
			 however, in this more general setting it is no longer generic and occurs only for certain values of the resetting point and of the coupling parameter. Indeed, expand 
			${\rm corr}(x,y)$
			 in Eq. ~\eqref{corr2} as a Taylor series in powers of 
			$r$. Retaining only the leading and the linear terms in 
			$r$,
			we obtain
				\begin{align}
					{\rm corr}(x,y) = - \frac{u (T_x + T_y)}{2 (1-u^2)} + \left(\frac{u (T_x + T_y)}{2 (1-u^2)^2} + \frac{2 X_0 Y_0 - u (X_0^2 + Y_0^2)}{4 (1 - u^2)}\right) \, r + O(r^2) \,.
					\end{align}
					Consider the case 
					$u<0$. Under this assumption, the occurrence of non-monotonic behavior requires the second term in the above expression to be positive. This takes place when 
					the coordinates of the resetting point, $T_x$ and $T_y$ and $u$ satisfy
					\begin{align}
						(1-u) X_0 Y_0 > - \frac{u (T_x + T_y)}{(1-u^2)} + \frac{u (X_0 - Y_0)^2}{2} \,,
						\end{align}
						which reduces for $X_0 = Y_0$ to a simpler inequality
						\begin{align}\label{eq:threshold}
							X_0^2  > - \frac{u (T_x  + T_y)}{(1+u)(1-u)^2} \,.
							\end{align}
							Therefore, for negative $u$, the $X_0$ has to be farther from the origin the larger $T_x+T_y$ is, in order for the non-monotonic behavior to take place. 
		\end{itemize}
		
\begin{figure}[h]
	\begin{center}
		\includegraphics[width=14cm]{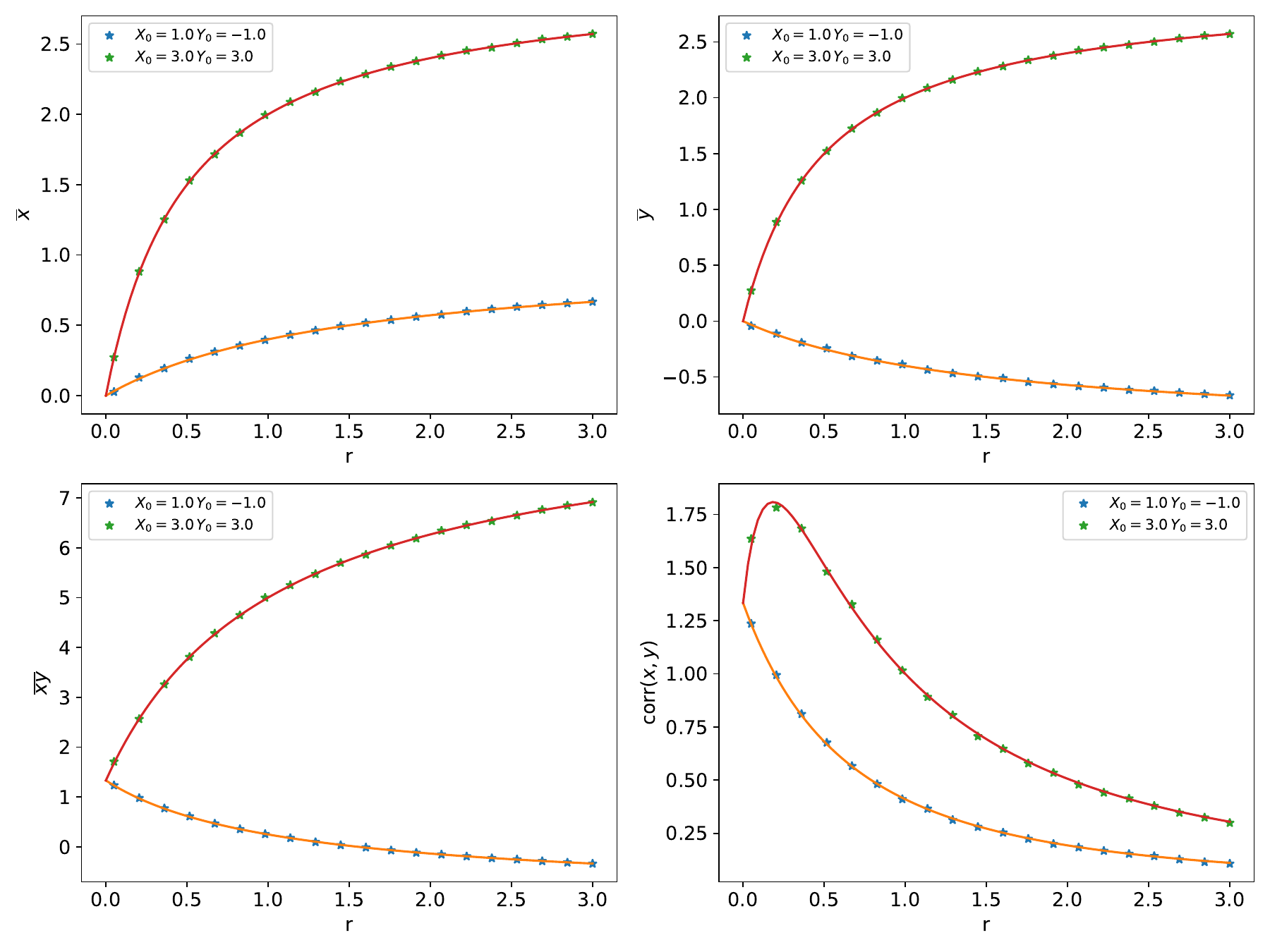}
	\end{center}
	\caption{Case A. The moments  $\overline{x}$ and $\overline{y}$, the cross-moment $\overline{xy}$ and the correlation function $\mathrm{corr}(x,y)$ for $u=-0.5$, $T_x=1$ and $T_y=3$, as functions of the resetting rate $r$ for two initial positions of the resetting point: $(X_0,Y_0)=(1,-1)$ and $(3,3)$. The solid lines represent our theoretical predictions in Eqs. \eqref{eq:mom}, \eqref{eq:mom1} and \eqref{corr2}, while the symbols here and henceforth denote numerical simulation results obtained by forward discretisation of the Langevin equations of motion, Eqs. \eqref{Lan} (see Appendix \ref{appendix:momentum}).}\label{fig:moments1}
\end{figure}

Figure ~\ref{fig:moments1} illustrates the behavior of the moments $\overline{x}$ and $\overline{y}$, the cross-moment  $\overline{xy}$ and the correlation function $\mathrm{corr}(x,y)$ as functions of the resetting rate $r$. The numerical results, obtained from simulations of the BG model with resetting to a fixed point $(X_0,Y_0)$, are compared with the exact analytical predictions given in Eqs.  ~\eqref{eq:mom} and ~\eqref{eq:mom1}, showing excellent agreement over the full range of parameters. Furthermore, we observe that in agreement with our preceding reasonings, the correlation function $\mathrm{corr}(x,y)$ can indeed be  either a monotonic or a non-monotonic function of $r$ attaining a maximal value at a certain finite resetting rate $r = r^*$, when the position of the resetting point obeys Eq.  \eqref{eq:threshold}.
The simulation results were obtained by averaging over $9\times 10^{4}$ trajectories, ensuring good statistical accuracy across the explored range of resetting rates.

	Further on, we find that the moments of the second order are given explicitly by
	\begin{align}\label{eq:mom2}
		\begin{split}
			\overline{x^2}  &= \frac{2 (2+r)^2 T_x - 4 u^2 \Delta_T + r (2 + r)^2 X_0^2 - 2 u r (2+r) X_0 Y_0+ 2 r u^2 (Y_0^2 - X_0^2)}{(2+r) \left((2 + r)^2 - 4 u^2 \right)} \,, \\
			\overline{y^2}  &= \frac{2 (2+r)^2 T_y + 4 u^2 \Delta_T + r (2 + r)^2 Y_0^2 - 2 u r (2+r) X_0 Y_0+ 2 r u^2 (X_0^2 - Y_0^2)}{(2+r) \left((2 + r)^2 - 4 u^2 \right)} \,,
		\end{split}
	\end{align}
	and show quite a complicated dependence on $r$, $T_x$, $T_y$ and $u$. As above, consider the dependence of the variance of one of the components, say, $x$, on the resetting rate. For simplicity, we start with the decoupled case $u = 0$, for which
	\begin{align}
		\begin{split}
			{\rm Var}(x) &=  \frac{2 T_x}{2 + r} + \frac{r X_0^2}{(2+r)(1+r)^2} \,.
			\end{split}
			\end{align}
We observe that for 
$u=0$
the variance of 
$x$
exhibits a non-monotonic dependence on the resetting rate 
$r$, which originates from the second term in the above expression. This non-monotonicity becomes increasingly pronounced as the resetting point 
$X_0$
moves farther away from the origin, reflecting the growing influence of the resetting-induced excursions. In contrast, the first term, which is proportional to 
$T_x$, decreases monotonically with 
$r$.
This behavior highlights the competition between two distinct contributions: the thermal fluctuations, encoded in the first term, which are progressively suppressed by frequent resetting, and the resetting-induced contribution, captured by the second term, which initially enhances fluctuations before being damped at large 
$r$. As a result of this interplay, the overall variance displays a characteristic non-monotonic dependence on the resetting rate.

For $u \neq 0$, the non-monotonic behavior of the variance as a function of $r$ takes place only if  some  conditions imposed on the coordinates of the resetting point, the  temperatures and $u$ are fulfilled. As above expanding the variance in the Taylor series in powers of $r$ and retaining only the constant and the linear terms, we have
\begin{align}
	\begin{split}
	{\rm Var}(x) &= \frac{2 T_x - u^2 \Delta_T}{2 (1 - u^2)} + \Bigg[\frac{4 T_x + 2 X_0^2 - 2 u  X_0Y_0 - u^2(X_0^2 - Y_0^2)}{4 (1-u^2)} \\
	&- \frac{(3 - u^2)(2 T_x - u^2 \Delta_T)}{4 (1 - u^2)^2} \Bigg] \, r + O(r^2) \,.
	\end{split}
	\end{align} 	
	Since the first term in the above expression is always positive and the variance vanishes as $r \to \infty$, a non-monotonic dependence of the variance on 
	$r$
	arises only if the contribution of the bracketed term is positive. This requirement leads to a rather involved inequality. However, the condition simplifies in the symmetric case 
	$X_0=Y_0$, where it reduces to
	\begin{align}\label{eq:threshold2}
		X_0^2 > \frac{2 (1+u^2) T_x - (3 - u^2) u^2 \Delta_T}{2 (1-u) (1-u^2)} \,.
		\end{align} 	
		
\begin{figure}[h]
		\begin{center}
			\includegraphics[width=14cm]{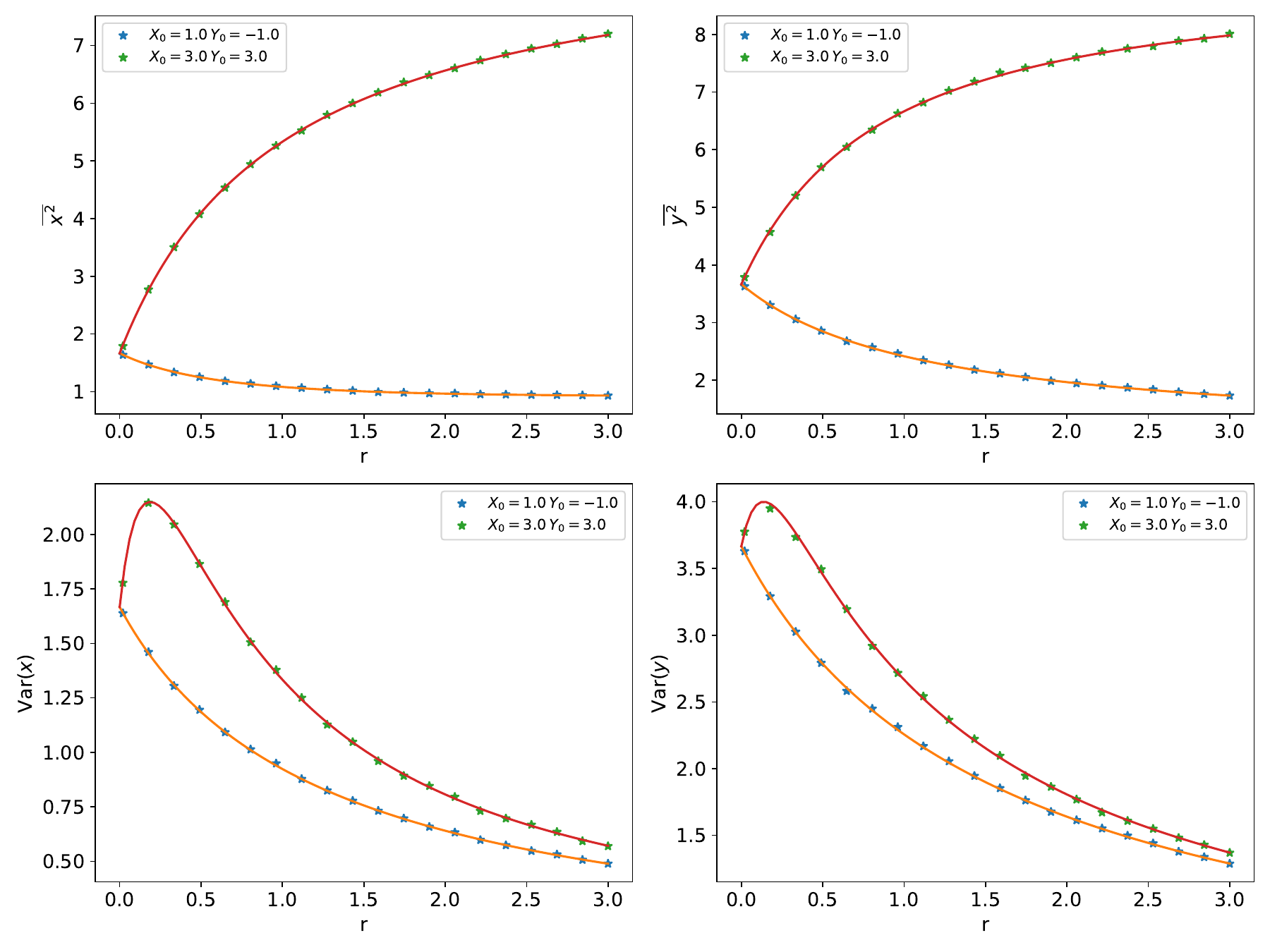}
		\end{center}
		\caption{Case A.
			Second moments $\overline{x^2}$, $\overline{y^2}$ and variances $\mathrm{Var}(x) = \overline{x^2}-\overline{x}^2$ and $\mathrm{Var}(y)=\overline{y^2}-\overline{y}^2$ versus the resetting rate $r$ for $T_x=1$, $T_y=3$, $u=-0.5$ and two positions of the resetting point: $(X_0,Y_0)=(1,-1)$ and $(3,3)$. The red solid curves are our analytical predictions  (see Eqs. \eqref{eq:mom} and \eqref{eq:mom2}), while symbols depict the results of numerical simulations. }\label{fig:moments2}
	\end{figure}

On the top of Fig. \ref{fig:moments2}, $\overline{x^2}$ and $\overline{y^2}$  are displayed as functions of the resetting rate $r$. The dots correspond to simulation  results and full curves depict the exact results, given in Eq. \eqref{eq:mom2} for two initial conditions: $(X_0,Y_0)=(1,-1)$ and $(3,3)$. When $r\rightarrow 0$, the second moments attain exactly the same form as their counterparts for the standard BG model, as they should. Conversely, when $r$ is large, the second moments depend strongly on the positions of the resetting point. On the bottom of Fig. \ref{fig:moments2}, we plot the variances ${\rm Var}(x)$  and ${\rm Var}(y)$ as functions of $r$. We observe that these properties 
indeed attain a maximum at some resetting rate when the inequality in Eq. \eqref{eq:threshold2} is fulfilled.
		
\subsection{Moment of inertia $I$}

From Eqs. \eqref{eq:mom2} we find the mean value of the specific (we set the mass $m=1$) moment of inertia $I = x^2+y^2$ in the presence of stochastic resetting to a fixed point $(X_0,Y_0)$. This mean value is given by,
for arbitrary values of parameters of the model,
\begin{align}\label{eq:I}
\overline{I} =  \frac{2 (2 + r) (T_x + T_y) + r (2 + r) X_0^2 - 4 u r X_0 Y_0 + r (2 + r) Y_0^2  }{(2 + r)^2 - 4 u^2} \,.
\end{align}
Further on, in order to determine the PDF  $P^{\rm st}_r(I|X_0,Y_0)$ of the specific 
moment of inertia, we proceed as follows: This PDF is formally defined as
\begin{align}
	P^{\rm st}_r(I|X_0,Y_0) = \int^{\infty}_{-\infty} \int^{\infty}_{-\infty}  dx \, dy \, \delta\left(x^2 + y^2 - I\right) \, P^{\rm st}_r(x,y|X_0,Y_0) \,,
	\end{align}
where $P^{\rm st}_r(x,y|X_0,Y_0)$ is given in Eq. \eqref{pr2}. Representing next the position PDF through its characteristic function in Eq. \eqref{Phi}, we get
\begin{align}
	\begin{split}
		\label{II}
		P^{\rm st}_r(I|X_0,Y_0) &= \frac{r}{4 \pi} \int^{\infty}_0 d\tau \exp(- r \tau) \int^{\infty}_{-\infty} \int^{\infty}_{-\infty} d\omega_1 \, d\omega_2 \, J_0\left(\sqrt{I (\omega_1^2 + \omega_2^2)}\right) \\& \times \exp\left( - i \omega_1 \tilde{X}_0 - i \omega_2  \tilde{Y}_0 - a \, \omega_1^2 + c  \, \omega_1 \omega_2 - b \, \omega_2^2  \right) \,,
	\end{split}
\end{align}
where $J_0(z)$ is the Bessel function \cite{AbramowitzStegun1964}, which 
can be conveniently expressed as  
	\begin{align}	
		\label{represent}
		J_{0}\left(\sqrt{I \left(\omega_1^2 + \omega_2^2\right)}\right) = \frac{1}{2   \pi} \int^{\pi}_{-\pi} d\theta \, \exp\left(i \, \sqrt{I} \, \omega_1 \, \cos(\theta) - i \, \sqrt{I} \, \omega_2 \, \sin(\theta)\right) \,,
	\end{align}
	such that the kernel depends on $\omega_1$ and $\omega_2$ in a factorized form. Inserting the above representation into Eq. \eqref{II} and  integrating over $\omega_1$ and $\omega_2$, we get
	\begin{align}
		\begin{split}
		\label{Pst}
		P^{\rm st}_r(I|X_0,Y_0) = \frac{r}{4 \pi} &\int^{\infty}_0 \frac{d\tau}{\sqrt{d}} \exp(- r \tau) \int^{\pi}_{-\pi} d\theta \, \exp\left(- \frac{b Z_x^2 - c Z_x Z_y + a Z_y^2}{d}\right) \,, \\
			Z_x&= \sqrt{I} \, \cos(\theta) - \tilde{X}_0 \,, \quad Z_y = \sqrt{I} \, \sin(\theta) + \tilde{Y}_0 \,.
		\end{split}
	\end{align}
\begin{figure}[h]
	\centering
	\includegraphics[width=14cm]{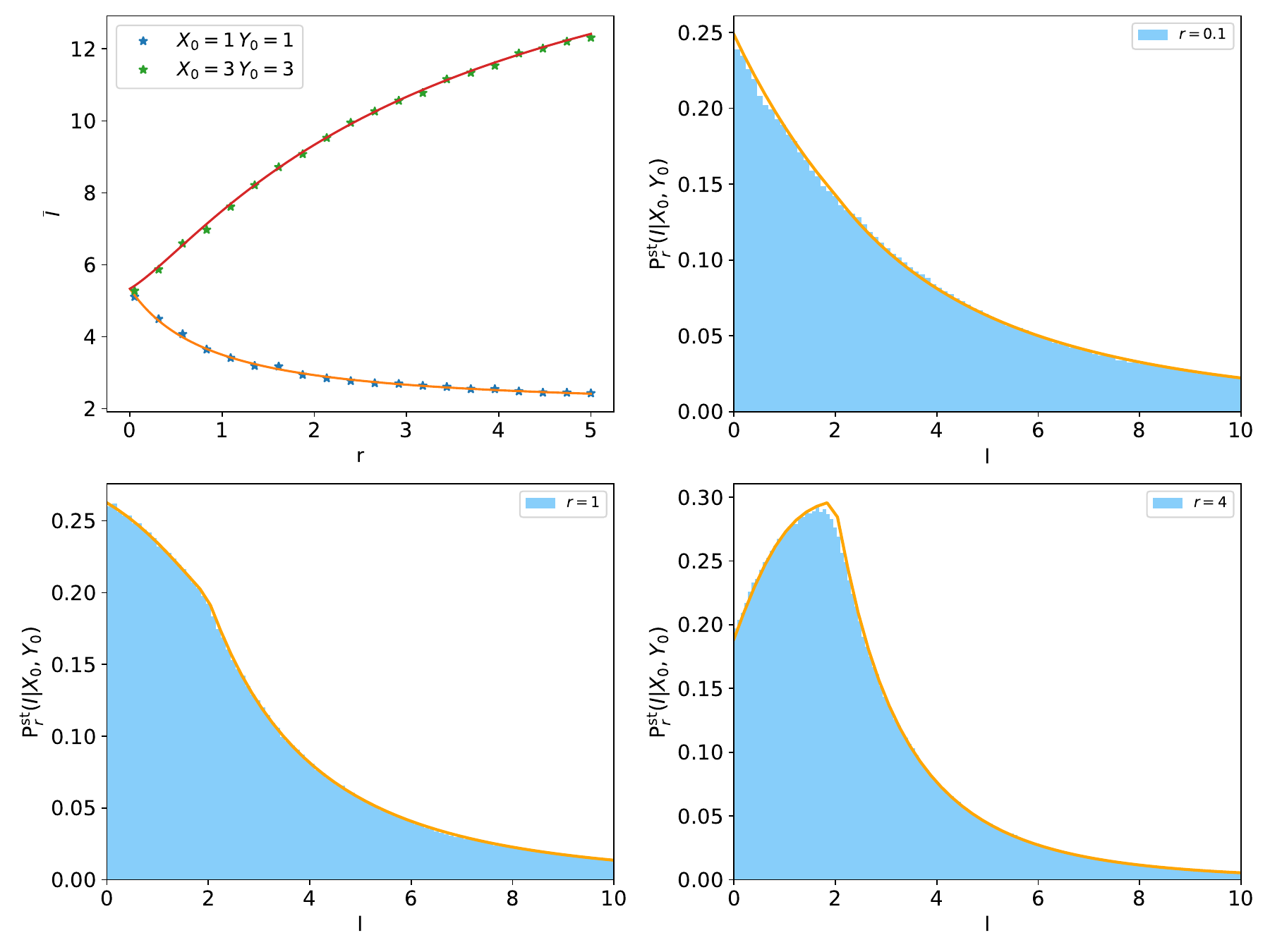}
	\caption{Case A. The moment of inertia for $T_x=1$,  $T_y=3$, $u=0.5$ and two positions of the resetting point: $(X_0,Y_0)=(1,1)$ and $(3,3)$. Top row, left panel: The mean moment of inertia $\overline{I}$ in Eq. \eqref{eq:I} as  function of $r$. Symbols depict the results of numerical simulations. Three other panels: the PDF $P^{\rm st}_r({\cal I}|X_0,Y_0)$ for $r=0.1$, $1$ and $4$ and $(X_0,Y_0)=(1,1)$. The histograms show the PDFs obtained via numerical simulations of Eqs. \eqref{Lan}, while the solid curves depict the exact result in Eq. \eqref{Pst}.}\label{fig:I}
\end{figure}
The top-left panel of Fig.~\ref{fig:I} shows the mean moment of inertia $\overline{I}$  as a function of the resetting rate $r$, for $T_x=1$, $T_y=3$ and $u=0.5$, and for two different positions of the resetting point: $(X_0,Y_0)=(1,1)$  and $(3,3)$. The solid curves depict the exact result in Eq.~\eqref{eq:I}, while the stars represent the simulation results. The three other panels show the PDF $P^{\rm st}_r(I|X_0,Y_0)$ for different values of the resetting rate $r$, with $(X_0,Y_0)=(1,1)$. The histograms depict the simulation results, while the solid curves are obtained by a numerical evaluation of the two-fold integral in Eq.~\eqref{Pst}.
For $r\ll1$, the PDF has an approximately exponential form, similar to that of the BG without resetting. At $r=1$, a qualitative change in the shape of the PDF becomes apparent: it is convex for small values of $I$ and becomes concave for $I > 2$, where $I=2$ corresponds to the moment of inertia at the resetting point. For $r>1$, the maximum of $P^{\rm st}_r(I|X_0,Y_0)$ shifts away from the origin toward the value of $I$ associated with the resetting point.

\subsection{The local probability current in the steady-state}

We next consider the so-called \textit{local} probability current 
$\vec{j} = (j_x,j_y)$ in the steady state  \cite{Mori2023}. By this one means the contribution to the stationary probability current arising solely from the continuous diffusive dynamics of the BG, excluding the non-local contributions associated with instantaneous resetting events. In the presence of resetting, the full probability current contains both a \emph{local} part, generated by the underlying Langevin dynamics, and a \emph{non-local} component corresponding to discontinuous relocations of the particle to the resetting point. The local current 
$\vec{j} = (j_x,j_y)$
therefore characterises the intrinsic circulation and transport properties of the system between successive resets, and provides a meaningful description of the steady-state flow structure in the configuration space.
The components of the local current are defined in a standard way as
\begin{equation}
	\begin{split}
		\label{j}
		j^{(X_0,Y_0)}_x &= - T_x \frac{\partial P^{\rm st}_r(x,y|X_0,Y_0)}{\partial x} - (x + u y) P^{\rm st}_r(x,y|X_0,Y_0) \,,\\
		j^{(X_0,Y_0)}_y &= - T_y \frac{\partial P^{\rm st}_r(x,y|X_0,Y_0)}{\partial y} -  (y + u x) P^{\rm st}_r(x,y|X_0,Y_0)
	\end{split}
\end{equation}
\begin{figure}[h]
	\centering
	\includegraphics[width=55mm]{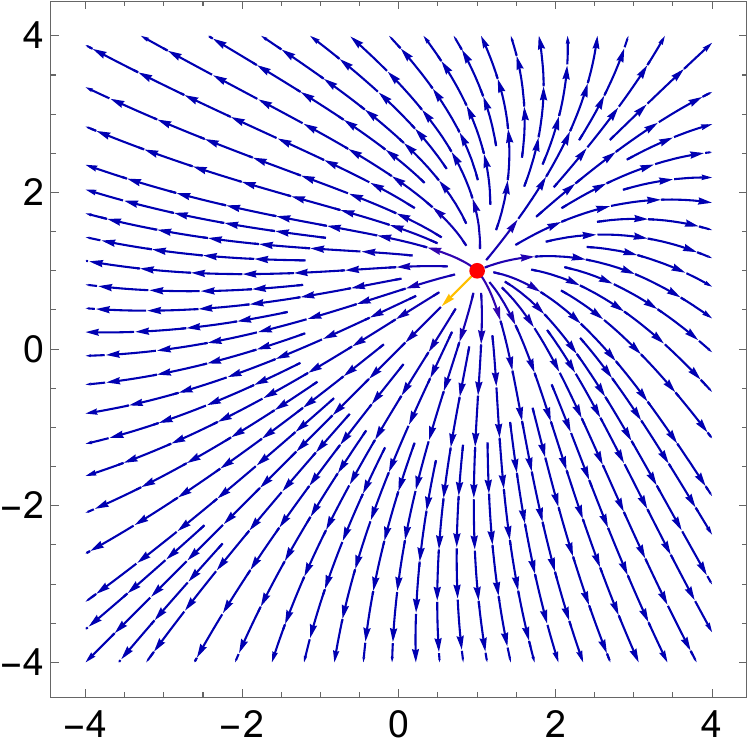}  
	\includegraphics[width=55mm]{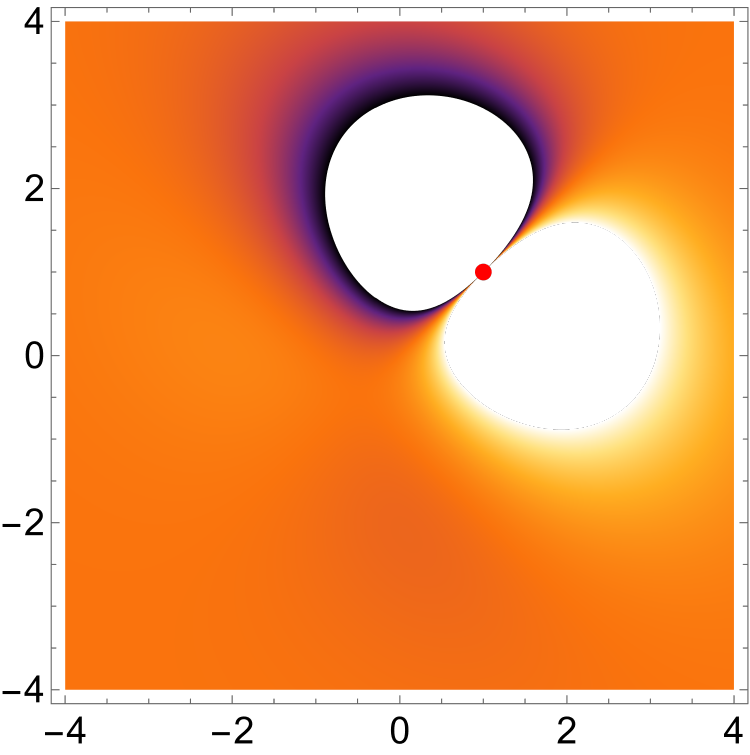}
	\includegraphics[width=13mm]{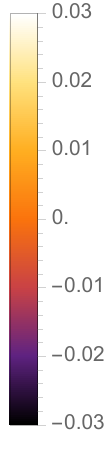}
	\caption{Case A.  Left panel: The stream plot of the probability current $\vec{j}^{(X_0,Y_0)}_r$ in Eq. \eqref{j} on the $(x,y)$-plane. The coupling parameter $u = 0.5$, 
	 	$X_0=Y_0 = 1$  (indicated by a filled red circle) and the temperatures $T_x = T_y = 4$ (no gyration).
		Right panel: Vorticity of 
		the probability current (see Eq.~\eqref{curl}) for the same values of parameters.}
	\label{fig:C} 
\end{figure}

Figure~\ref{fig:C} (left panel) shows that stochastic resetting modifies the stationary probability current in a more profound way than merely changing its magnitude. To disentangle the contribution generated by resetting from the intrinsic BG current, we choose in Fig.~\ref{fig:C} the case of equal temperatures, $T_x=T_y=4$, in which 
the local probability current vanishes for the standard BG model. The nonzero current displayed in the left panel of Fig.~ \ref{fig:C} is therefore entirely induced by the resetting dynamics.

A particularly striking feature is that this resetting-induced current is not simply directed toward the resetting point. Recall that $\vec{j}_r$ describes only the continuous part of the dynamics between resetting events; the instantaneous relocation of probability to $(X_0,Y_0)$ constitutes a separate, non-local contribution to the full probability current. The local dynamics must consequently redistribute the probability continuously away from the probability injected at the resetting point. This produces the strongly distorted flow pattern seen in Fig.~\ref{fig:C}. Thus, even though the underlying Langevin dynamics at $T_x=T_y$ is an equilibrium process, its combination with resetting generates a genuine non-equilibrium stationary state. This observation makes explicit that stochastic resetting constitutes an independent source of time-reversal-symmetry breaking, rather than merely modifying the non-equilibrium state already generated by $T_x\neq T_y$. The distinction between the local diffusive current and the non-local resetting contribution is made explicitly in the definition preceding Eq.~\eqref{j}.

The right panel of Fig.~\ref{fig:C} provides an even stronger manifestation of this effect. The vorticity is nonzero and changes sign in different regions of the $(x,y)$-plane, despite the equality of the two temperatures. Resetting can therefore generate a spatially-structured \emph{rotational} component of the local probability flow even when the conventional BG mechanism is absent. The positive and negative regions of vorticity indicate that the resetting-induced flow cannot be characterised by a single global sense of rotation; instead, resetting creates counter-rotating regions whose geometry is controlled by the position of the resetting point relative to the confining potential.

\begin{figure}[h]
	\centering
	\includegraphics[width=50mm]{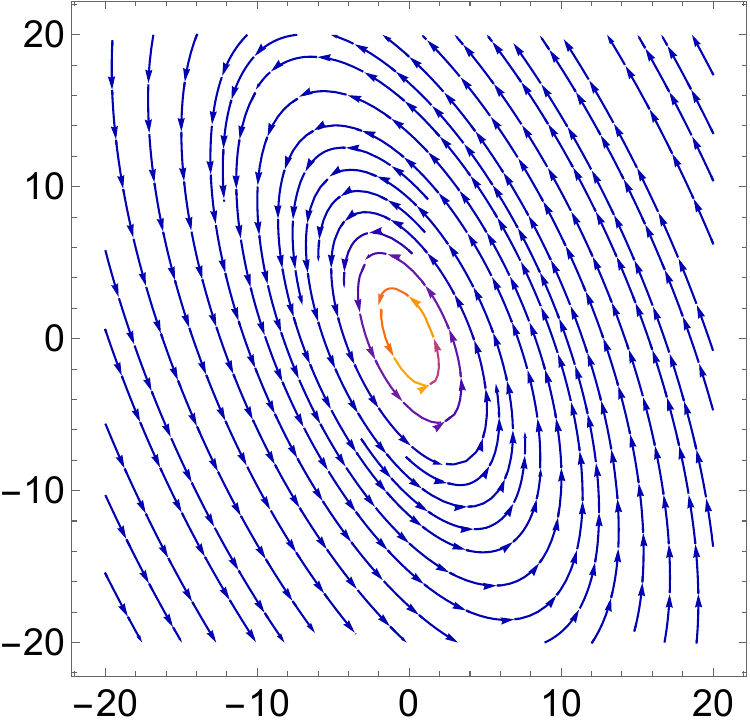}  
	\includegraphics[width=50mm]{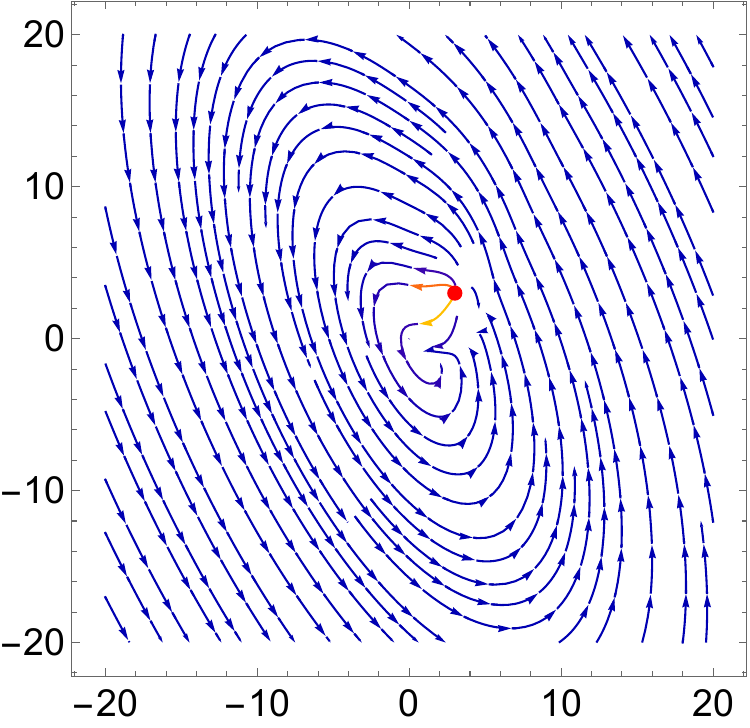}
	\includegraphics[width=50mm]{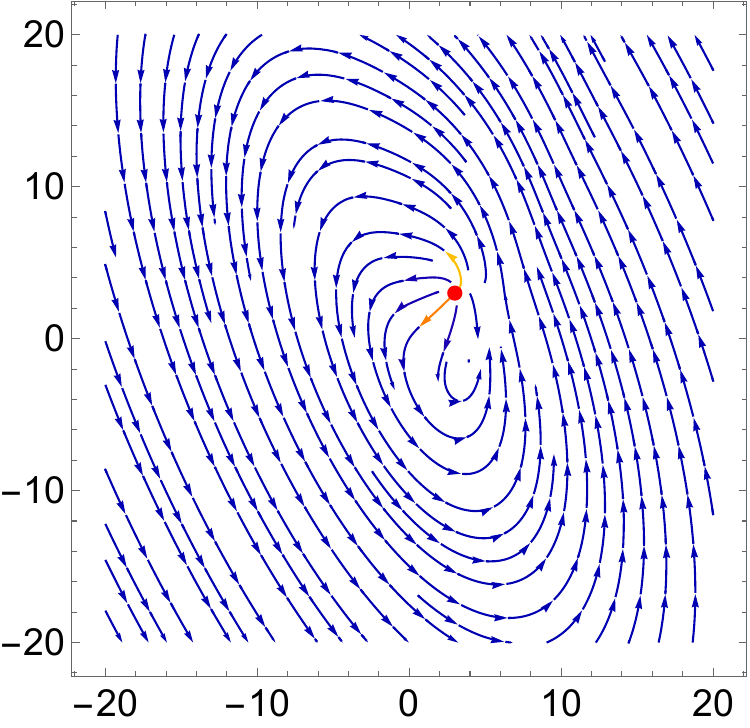}
	\caption{Case A. The stream plot of the probability current $\vec{j}^{(X_0,Y_0)}_r$ in Eq. \eqref{j} on the $(x,y)$-plane 
		for $u=0.5$,  $X_0 = Y_0 = 3$ (filled red circle), $T_x = 1$ and $T_y = 4$. Left panel: $r = 0$ (no resetting). Middle panel: $r = 0.1$. Right panel: $r = 1$.}
	\label{fig:2} 
\end{figure}

Figure~ \ref{fig:2} demonstrates the competition between this resetting-induced current and the intrinsic rotational current of the BG model. For $r=0$ (left panel), the current follows the familiar closed elliptical streamlines centered around the origin, reflecting the anisotropy generated by $T_x\neq T_y$ in the coupled harmonic potential. Upon introducing a weak resetting rate, $r=0.1$ (middle panel), the elliptical circulation survives but its spatial symmetry is broken. The streamlines become distorted toward the resetting point $(X_0,Y_0)=(3,3)$, showing that even relatively rare resets reorganise the stationary probability transport over a substantial region of configuration space. For $r=1$ (right panel), this distortion is much stronger: the resetting point becomes a prominent organising center of the flow and the simple nested elliptical structure characteristic of the standard Brownian gyrator is largely lost.

These observations reveal a crossover between two distinct non-equilibrium mechanisms. At small $r$, the stationary current is dominated by the temperature-induced gyration and resetting acts primarily as a perturbation of the elliptical circulation. As $r$ increases, the renewal dynamics associated with repeated returns to $(X_0,Y_0)$ becomes increasingly important, breaking the approximate spatial symmetry of the BG current and concentrating the flow around the resetting location. In the large-$r$ regime, trajectories have progressively less time to explore the harmonic potential between successive resetting events, so that the local current is expected to become increasingly localised in the neighbourhood of the resetting point. This interpretation is also consistent with the position PDF discussed earlier in the manuscript: increasing $r$ transfers the dominant probability weight from the neighbourhood of the origin toward the resetting point, eventually producing a singular peak there.

It is important, however, not to interpret the deformation of the streamlines as a simple suppression of rotational dynamics. Resetting introduces its own rotational structure, as demonstrated independently by Fig.~\ref{fig:C}. The current observed in Fig.~ \ref{fig:2} is therefore the result of a superposition and competition between the intrinsic BG circulation and a resetting-induced non-equilibrium flow. Depending on position, these two contributions may reinforce or oppose one another. This spatial competition naturally anticipates the more complex vorticity patterns discussed below.

\subsection{Curl (vorticity) of the probability current. Spinning of particles}

The curl (vorticity) of $\vec{j}$ is defined by
\begin{align}
	\label{curl}
	{\rm curl} \vec{j}^{(X_0,Y_0)} = \left(\frac{\partial j_y}{\partial x} - \frac{\partial j_x}{\partial y}\right) \vec{e}_z
\end{align}
where $ \vec{e}_z$ is a unit vector directed along the $z$-axis. If this property is not identically equal to zero, this will mean that a finite size particle will perform also a spinning around its center of mass with the direction and the amplitude being defined by the sign and the value of the terms in brackets (see e.g. \cite{Dotsenko2022a,Dotsenko2023}). 

\begin{figure}[h]
	\centering
	\includegraphics[width=42mm]{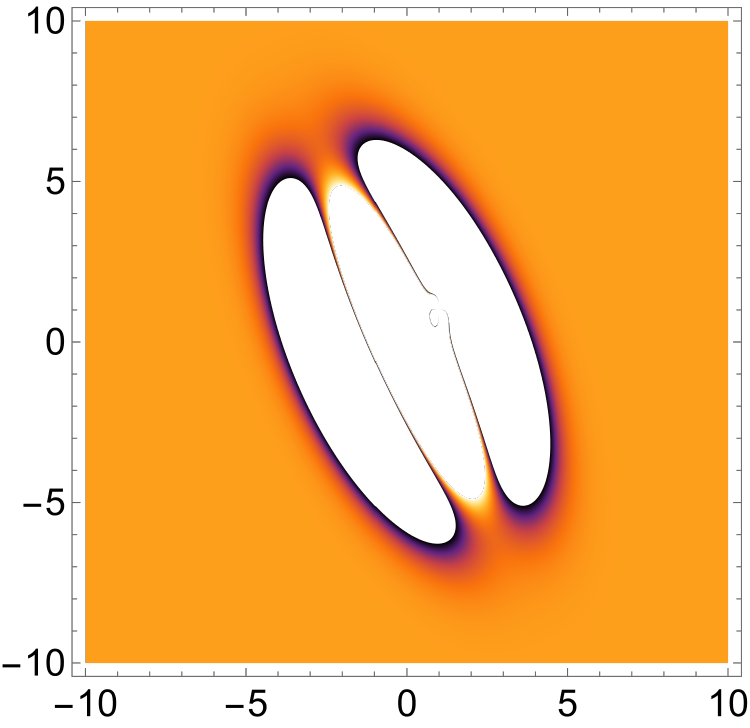}  
	\includegraphics[width=42mm]{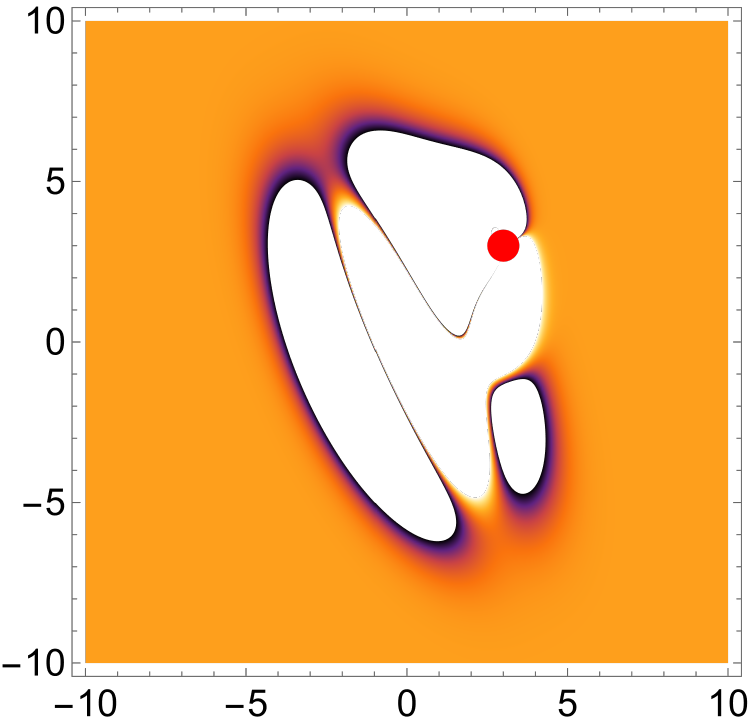}	\includegraphics[width=42mm]{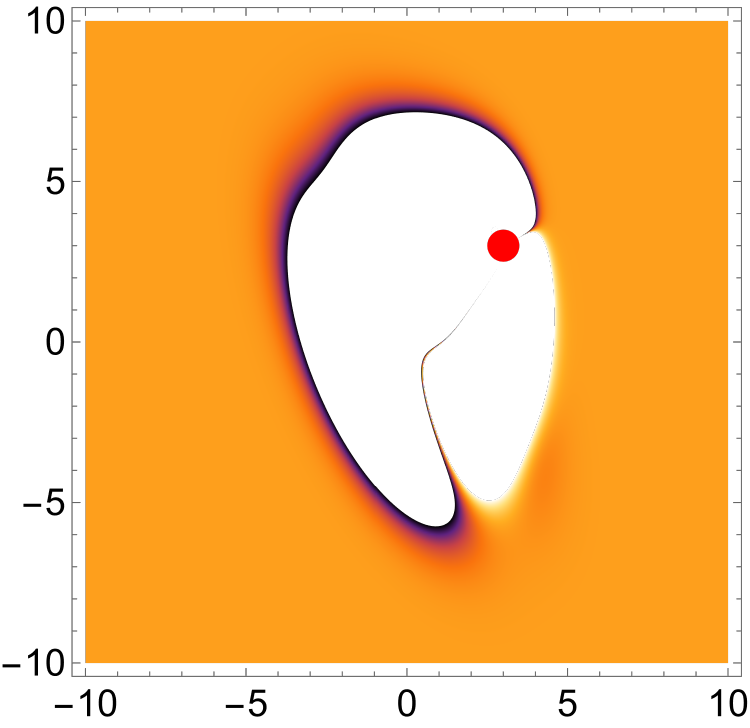}
	\includegraphics[width=10.5mm]{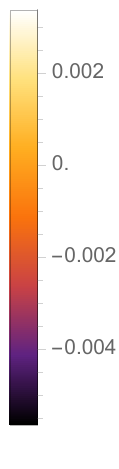}
	\caption{Case A. The amplitude of the vorticity of $\vec{j}^{(X_0,Y_0)}$ (see Eq.  \eqref{curl}) on the $(x,y)$-plane 
		for $u=0.5$, $X_0 = Y_0 = 3$ (red filled circle), $T_x = 1$ and $T_y = 4$. Left panel: $r = 0$ (no resetting). Middle panel: $r = 0.1$. Right panel: $r = 1$.}
	\label{fig:3} 
\end{figure}

Figure~ \ref{fig:3} shows how stochastic resetting reorganises the spatial structure of the vorticity field. In the absence of resetting ($r=0$, left panel), the vorticity originates exclusively from the conventional BG mechanism, namely the combined effect of the coupling $u$ and the temperature anisotropy $T_x\neq T_y$. The field displays a regular structure centered around the minimum of the confining potential, reflecting the elliptical circulation of the stationary probability current observed in the corresponding $r=0$ panel of Fig.~\ref{fig:2}. The sign of the vorticity determines the local sense of rotation, while its magnitude measures the local rotational strength of the probability flow; for a finite-size particle, it may consequently be associated with a local tendency to spin about its center of mass, as expressed by Eq.~\ref{curl}.

The introduction of even a relatively weak resetting rate, $r=0.1$ (middle panel), qualitatively modifies this structure. The vorticity is no longer organized solely around the origin but becomes spatially asymmetric as a consequence of the off-center resetting point $(X_0,Y_0)=(3,3)$. This deformation parallels the distortion of the probability-current streamlines seen in Fig.~ \ref{fig:2}. It demonstrates that resetting does not simply rescale the vorticity already present in the BG  model: it generates an additional, spatially inhomogeneous contribution whose symmetry is determined by the resetting position. The resulting vorticity field therefore contains information that cannot be inferred from a global measure of rotation alone.

At the larger resetting rate $r=1$ (right panel), the restructuring becomes still more pronounced. The characteristic vorticity pattern of the unperturbed gyrator is strongly deformed and the influence of the resetting point becomes increasingly localised and visible. This behavior has a natural interpretation in terms of the renewal representation of the stationary state. Increasing $r$ gives progressively larger statistical weight to trajectories observed shortly after their most recent resetting  event. The system therefore has less time to relax from $(X_0,Y_0)$ toward the region favored by the harmonic confinement before another resetting event occurs. Consequently, the local gradients of the stationary probability density, and hence the local current and its curl, become increasingly controlled by the resetting point rather than solely by the temperature-induced gyration.

Figure~ \ref{fig:C} demonstrates that resetting itself can generate a nonzero and spatially structured vorticity even for $T_x=T_y$, when the conventional thermal mechanism responsible for gyration is absent. Figure~ \ref{fig:3} should therefore be viewed as displaying the competition between two distinct sources of rotational probability flow: the intrinsic gyration generated by $T_x\neq T_y$, and the additional non-equilibrium circulation generated by stochastic resetting. Since these two contributions have different spatial symmetries -- the former being associated with the origin and the anisotropic stationary state, and the latter with the displaced resetting point -- their superposition produces the increasingly complex vorticity patterns observed as $r$ grows.

The occurrence of regions with different signs of the vorticity is particularly noteworthy. It implies that the local sense of rotation is position dependent: stochastic resetting can enhance the original gyration in some regions while opposing it in others. Hence a single global quantity, such as the mean angular momentum or mean angular velocity considered below, necessarily averages over a much richer spatial organisation of the rotational motion. The vorticity field thus provides complementary information by resolving where in configuration space the rotational probability flow is generated, enhanced, suppressed, or even reversed.

\subsection{Angular momentum for resetting to a fixed point}

We first consider the specific angular momentum, defined with the orientation $
 \overrightarrow{\cal L}=\vec r\times\vec v $, where the symbol $\times$ denotes the vector product, 
so that its only non-zero component is
\begin{equation}
	\label{L}
 {\cal L}=x \dot{y}- y \dot{x} .
\end{equation}
For the stationary state, the mean angular momentum can be expressed in terms of the local probability current in Eq.~\eqref{j} as
\begin{align}
	\label{eq:angular}
	\begin{split}
\overline{\cal L} &=\int_{-\infty}^{\infty}\int_{-\infty}^{\infty} dx\,dy\,
\Bigg[u(y^2-x^2) P^{\rm st}_r(x,y|X_0,Y_0) \\
&+T_x y\frac{\partial P^{\rm st}_r(x,y|X_0,Y_0)}{\partial x}
-T_y x\frac{\partial P^{\rm st}_r(x,y|X_0,Y_0)}{\partial y}\Bigg] \,.
\end{split}
\end{align}
Assuming that the stationary density decays sufficiently rapidly at infinity, integration by parts eliminates both derivative terms  (see Appendix \ref{appendix:momentum} for more details). Substitution of the renewal representation then yields
\begin{equation}\label{eq:angular2}
\overline{\cal L} =r u \int_0^\infty d\tau\,e^{-r \tau}\,
 \left(\widetilde{Y}_0^2-\widetilde{X}_0^2+2(b -a)\right).
\end{equation}
Using Eqs.~\eqref{tilde} and \eqref{coeft}, the remaining time integral gives the exact result
\begin{equation}\label{eq:Lmean}
\overline{\cal L} =\frac{u}{2+r}
\left[r(Y_0^2-X_0^2)+2(T_y-T_x)\right].
\end{equation}
An alternative derivation of Eq. \eqref{eq:Lmean} based on the forward discretization of the Langevin equations \eqref{Lan} is presented in Appendix ~\ref{appendix:momentum} (see also \cite{Viot2024}).

Equation~\eqref{eq:Lmean} separates the contributions of the two non-equilibrium mechanisms and gives a transparent illustration of their competition. The contribution $2(T_y-T_x)$ originates from the presence of two unequal temperatures which is responsible for the conventional BG rotation, whereas the term $r(Y_0^2-X_0^2)$ is generated by resetting and reflects the spatial anisotropy of the resetting position. For $r\to 0$, one recovers a known result $\overline{\cal L}= u (T_y-T_x)$ (see, e.g., \cite{Viot2024}). 
	As the resetting rate $r$ increases, the mean angular momentum approaches its limiting value $\overline{\cal L}=u(Y_0^2-X_0^2)$.  For $u=0$, $\overline{\cal L}$ vanishes: resetting alone does not produce a nonzero global mean angular momentum in this uncoupled model, even though it can generate nontrivial local currents.

\begin{figure}[H]
 \centering
  \includegraphics[width=15cm]{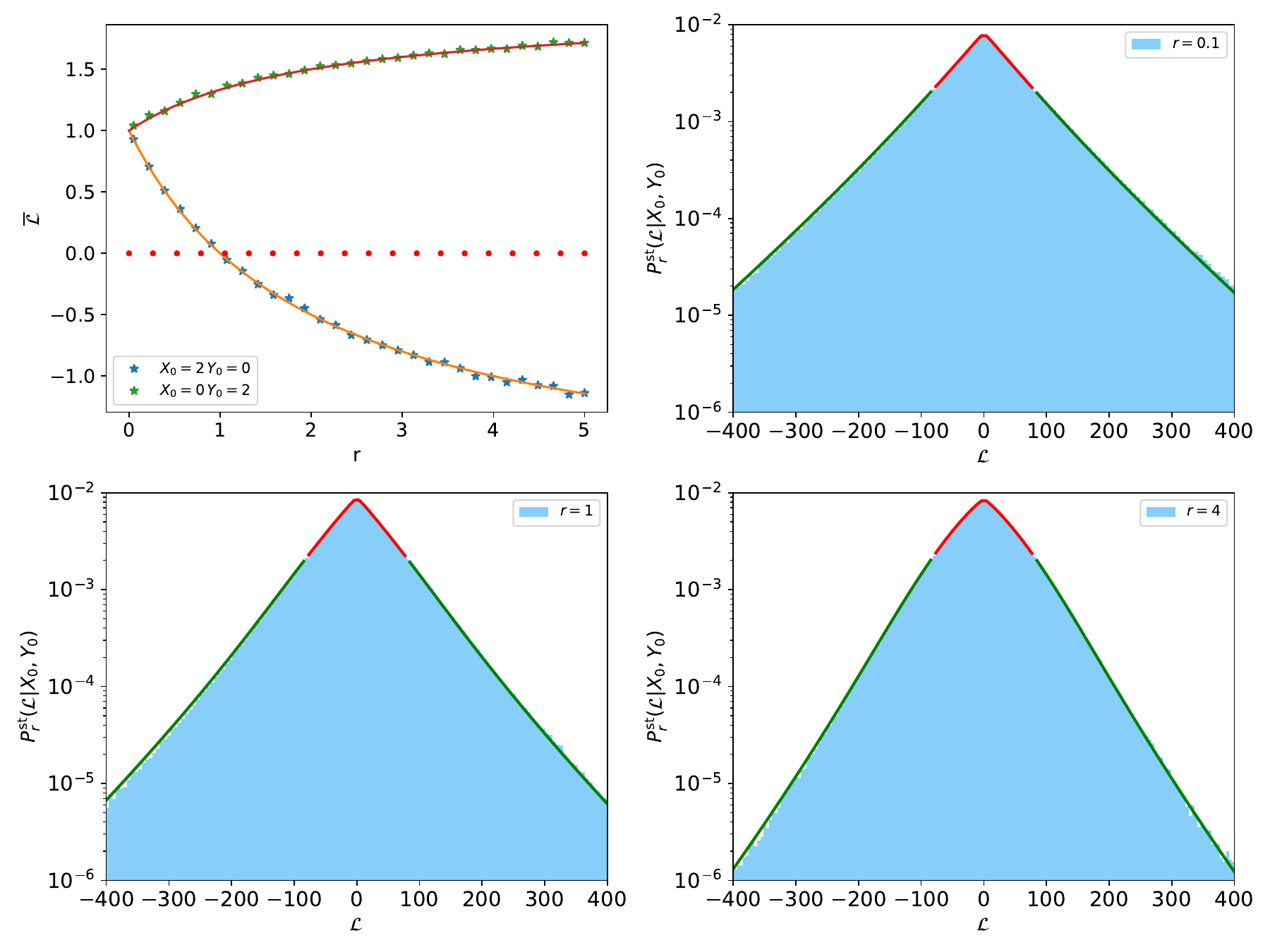}
 \caption{Case A. The specific angular momentum ${\cal L}$ for $T_x=1$, $T_y=3$  and $u=0.5$.  Top-left panel: $\overline{\cal L}$ versus $r$ for two positions of the resetting point: $(X_0,Y_0)=(2,0)$ and $(0,2)$. Solid curves are our analytical prediction in Eq. \eqref{eq:Lmean}, while symbols present the results of numerical simulations .  The horizontal dotted line serves as a guide to the eye, indicating zero mean angular momentum. Three other panels show the PDF $P^{\rm st}_r({\cal L}|X_0,Y_0)$ of stationary angular momentum  for the resetting rates indicated in the panels and $(X_0,Y_0)=(2,0)$. The histograms are the PDFs obtained via numerical simulations. The red curves are the results based on the numerical integration of Eq. \eqref{eq:PLpolar}, while  the green curves are exponential fits of the tails in Eq. \eqref{eq:Ltails}.}\label{fig:L}
\end{figure}

Remarkably, when the two contributions have opposite signs, their competition can reverse the sign of the mean angular momentum as the resetting rate increases. This is a reversal of the global mean rotational observable, not necessarily a reversal of the local current at every position. Setting $\overline{\cal L}=0$ gives for the critical resetting rate $r_{\rm inv}$
\begin{equation}
	r_{\rm inv}=-\frac{2(T_y-T_x)}{Y_0^2-X_0^2}
	=-\frac{2(T_x-T_y)}{X_0^2-Y_0^2} .
	\label{eq:rinv}
\end{equation}
Note that the reversal of the sign takes place only when $r_{\rm inv}$ is positive. When the right-hand-side of the above expression is negative, no inversion takes place.

The upper-left panel of Fig.~\ref{fig:L} illustrates this distinction for two resetting positions at the same distance from the origin. For $(X_0,Y_0)=(2,0)$, the resetting term opposes the thermal contribution and Eq.~\eqref{eq:rinv} predicts that $\overline{\cal L}$ crosses zero at $r = r_{\rm inv}=1$. For $(X_0,Y_0)=(0,2)$, the two terms have the same signs and therefore reinforce each other; the formal crossing is at $r=-1$ and is therefore inaccessible. Thus changing only the orientation of the resetting position can switch between a reinforcement of the mean rotation to its full reversal.

The mean alone does not describe the fluctuations around these trends. The full PDF $P^{\rm st}_r({\cal L}|X_0,Y_0)$ admits a three-fold integral representation, derived in Appendix~\ref{appendix:momentum} (red curves in Fig. ~\ref{fig:L}). The remaining panels of Fig.~\ref{fig:L} show distributions with support on both signs of $\cal L$, even when the mean has a preferred sign. On the displayed logarithmic scale, the far tails are approximately linear, suggesting the exponential forms
\begin{equation}
P^{\rm st}_r({\cal L}|X_0,Y_0)\sim
\begin{cases}
 A_-\exp(\lambda_-{\cal L}), & {\cal L}\to-\infty,\\[1mm]
 A_+\exp(-\lambda_+{\cal L}), & {\cal L}\to \infty,
\end{cases}
\label{eq:Ltails}
\end{equation}
with $\lambda_\pm>0$ (see Eq.\eqref{eq:lambdaLpm} for an explicit expression) The unequal positive- and negative-tail slopes quantify an asymmetry of large rotational fluctuations; the sign of the mean, however, is determined by the entire distribution and should not be inferred from the tails alone. Comparing the panels at different resetting rates illustrates that resetting changes the central weight and the relative probabilities of positive and negative angular momentum, while the tails remain approximately exponential. A useful empirical measure of the tail asymmetry is
\begin{equation}
 \ln\frac{P^{\rm st}_r({\cal L}|X_0,Y_0)}{P^{\rm st}_r(-{\cal L}|X_0,Y_0)}
 \simeq (\lambda_- - \lambda_+){\cal L}+\mathrm{const.},
 \label{eq:Lasym}
\end{equation}
within the range where the exponential fits apply. This relation characterizes the observed tails only; it is not assumed to constitute a fluctuation theorem. The comparison between the mean and the PDF makes clear why a change in the sign of $\overline{\cal L}$ need not imply that fluctuations of the opposite sign disappear.

\subsection{Angular velocity for resetting to a fixed point}

We define the instantaneous specific angular velocity, with the same orientation as the angular momentum ${\cal L}$, by
\begin{equation}\label{eq:Wdef}
 {\cal W}=\frac{x \dot{y}- y\dot x}{x^2+y^2} = \frac{{\cal L}}{I} ,
\end{equation}
where $I$ is the angular momentum.
Because the Langevin trajectories are not differentiable, Eq.~\eqref{eq:Wdef} is understood in the forward (It{\^o}) discretisation specified in Appendix~\ref{appendix:Pdew}.  Averaging first over the noise increments gives the compact exact relation
\begin{equation}\label{eq:angular4}
\overline{\cal W}
=u\int_{-\infty}^{\infty} \int_{-\infty}^{\infty} dx\,dy\,\frac{y^2-x^2}{x^2+y^2}
P^{\rm st}_r(x,y|X_0,Y_0).
\end{equation}
Thus, in contrast to the mean angular momentum, the mean angular velocity depends on the full stationary position distribution and not only on its second moments.  An explicit integral representation suitable for numerical evaluation is derived in Appendix~\ref{appendix:Pdew}.

\begin{figure}[h]
 \centering
\includegraphics[width=7cm]{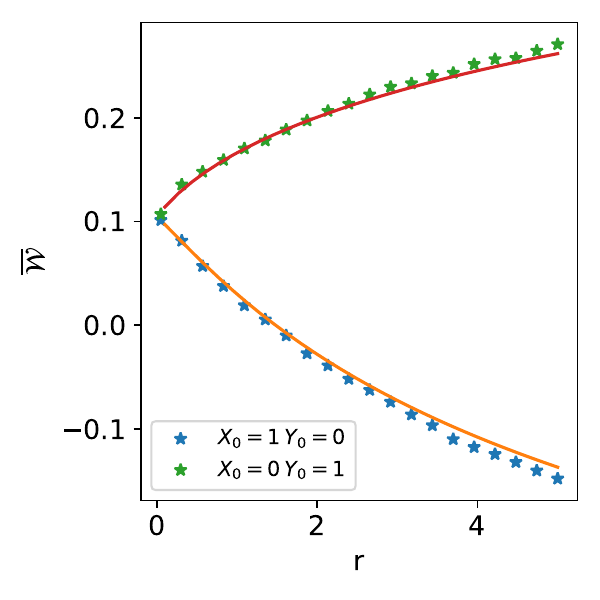}\includegraphics[width=7cm]{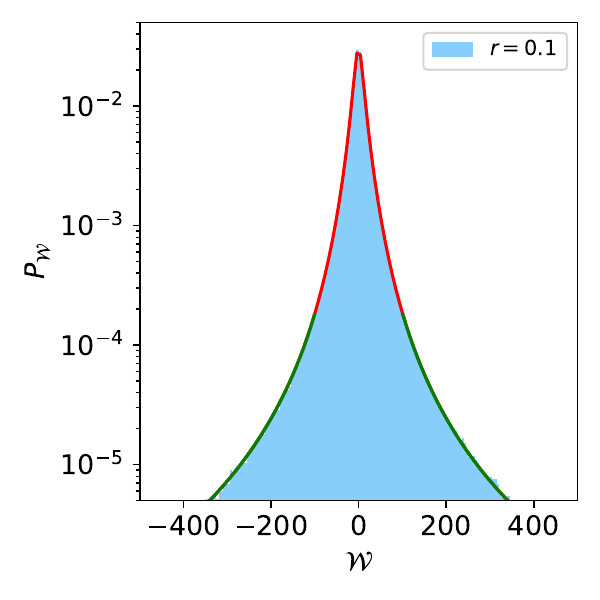}
\includegraphics[width=7cm]{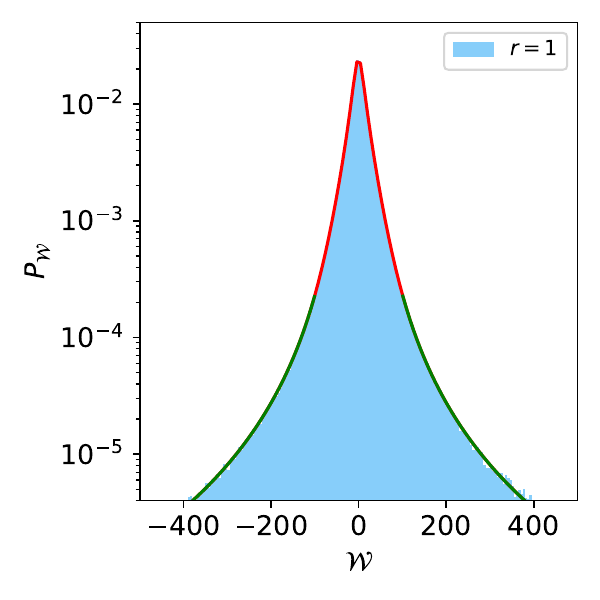}
\includegraphics[width=7cm]{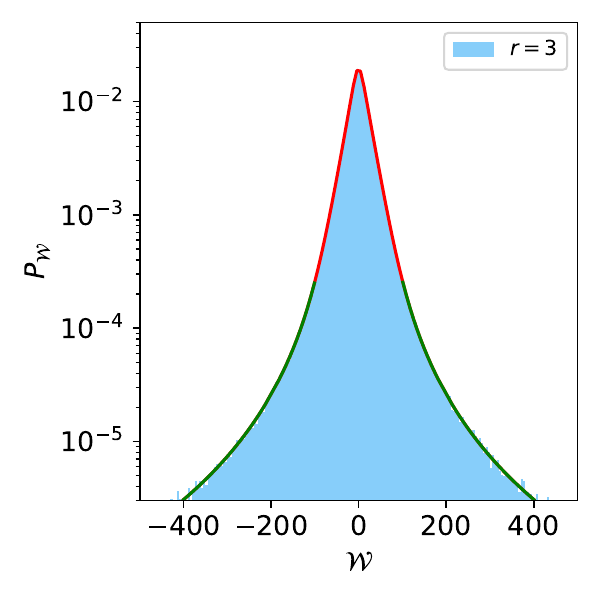}
 \caption{Case A. The specific angular velocity ${\cal W}$ for $T_x=1$, $T_y=3$ and $u=0.5$.
 	Top-left panel: $\overline{\cal W}$ versus $r$  for two resetting points: $(X_0,Y_0)=(1,0)$ and $(0,1)$. The solid curves show $\overline{\cal W}$ obtained via a numerical evaluation of the two-fold integral in Eq. \eqref{eq:angular4},
 		 while symbols present the results of numerical simulations. 
 	Three other panels depict the PDF $P_{\cal W} = P_r^{\rm st}({\cal W}|X_0,Y_0)$ for the resetting rates indicated in the insets and $(X_0,Y_0)=(1,0)$. The histograms are the results of numerical simulations based on Eqs. \eqref{Lan}. The red curves are obtained by a numerical integration of Eq. \eqref{eq:PWpolarSimple}, while the green curves are the fits of the PDF's power-law tails in Eq.~\eqref{eq:Wtail}. }
 \label{fig:W}
\end{figure}

A simple analytical result can nevertheless be obtained for the tails of the angular-velocity PDF.  For a finite observation step $\delta t$ and a resetting point different from the origin, the stationary position density is regular at the origin.  Large values of $|{\cal W}|$ are then generated predominantly by trajectories passing very close to the origin, where division by $x^2+y^2$ amplifies the instantaneous noise.  This gives (see Appendix~\ref{appendix:Pdew})
\begin{equation}\label{eq:Wtail}
	P_r^{\rm st}({\cal W}|X_0,Y_0)
	\underset{|{\cal W}|\to\infty}{\simeq}
	\frac{\pi(T_x+T_y)}{\delta t} 
	\,\frac{P_r^{\rm st}(0,0|X_0,Y_0)}{|{\cal W}|^3},
\end{equation}
For a resetting point away from the origin, $P_r^{\rm st}(0,0|X_0,Y_0)$ remains finite meaning that resetting changes the frequency of near-origin passages but does not exclude them completely. Consequently, in the presence of resetting 
to a fixed point the algebraic power-law tails of the angular-velocity PDF, which appear in the standard BG model \cite{Viot2024}, 
remain unchanged, although their amplitude acquires a dependence on the resetting rate and the location of the resetting point. The first absolute moment of ${\cal W}$ is therefore finite, while its second and higher moments diverge in the finite-$\delta t$ model. Eliminating this singular tail would require a mechanism that suppresses the probability density in a neighbourhood of the origin, rather than simply interrupting trajectories through resetting. The short-distance argument and its range of validity are given in Appendix~\ref{appendix:Pdew}.

The upper-left panel of Fig.~\ref{fig:W} compares the dependence of $\overline{\cal W}$ on the resetting rate for points $(X_0,Y_0)=(1,0)$ and $(0,1)$. Unlike the simple second-moment formula for $\overline{\cal L}$, these curves probe how resetting redistributes probability at different distances and orientations relative to the origin.
For the resetting point $(X_0,Y_0)=(1,0)$, the mean angular velocity $\overline{\cal W}$ changes sign at a critical value of the resetting rate $r$. It is positive for small $r$ and becomes negative above this critical value. This behavior is consistent with that of the mean angular momentum shown in Fig. \ref{fig:L}. In contrast, for $(X_0,Y_0)=(0,1)$, $\overline{\cal W}$ remains positive for all values of $r$, as does the mean angular momentum.
The three remaining panels of Fig. \ref{fig:W} show the full angular-velocity PDFs for increasing $r$. While resetting evidently reshapes the central parts of the PDFs and the relative weights of the two sides of the distribution, we also observe that the far tails are indeed algebraic with the exponent $3$, which confirms our prediction in Eq. \eqref{eq:Wtail}.

\section{Case B: Resetting to a circle $X_0^2 + Y_0^2 = R_0^2$}\label{sec:resetBG2}

Consider next the case B when the BG undergoes a stochastic 
 resetting to a random point on a circle  $X_0^2 + Y_0^2 = R_0^2$ centered around the origin. The integral representations of the position PDF $P_r^{\rm st}(x,y|R_0)$ and the corresponding Fourier transform $\Phi_r^{\rm st}(\omega_1,\omega_2|R_0)$ can be readily obtained by averaging 
 	Eqs. \eqref{pr2} and \eqref{Phi} 
over positions of the resetting point $(X_0,Y_0)$ uniformly distributed along the circumference of the circle. This yields	
\begin{align}
\begin{split}
\label{pcircle}
&P_r^{\rm st}(x,y|R_0)  = \langle P_r^{\rm st}(x,y|X_0, Y_0) \rangle_{\rm circle} \\
&= \frac{r}{(2 \pi)^2} \int^{\infty}_0 d\tau \, e^{- r \tau} \int^{\infty}_{-\infty} \int^{\infty}_{-\infty} d\omega_1 \, d\omega_2 \, 
\exp\left(i \omega_1 x  + i \omega_2 y  - a \, \omega_1^2 + c  \, \omega_1 \omega_2 - b \, \omega_2^2  \right) \\
& \qquad \times J_0\left(\ell \left(\omega_1^2 + \omega_2^2 - 2 \tanh(2 u \tau) \omega_1 \omega_2\right)^{1/2}\right) \,, \\
&\Phi_r^{\rm st}(\omega_1,\omega_2|R_0)  = \langle \Phi_r^{\rm st}(\omega_1,\omega_2|X_0, Y_0) \rangle_{\rm circle}\\
&=r  \int^{\infty}_0 d\tau \, e^{- r \tau}  \, J_0\left(\ell  \left(\omega_1^2 + \omega_2^2 - 2 \tanh(2 u \tau) \omega_1 \omega_2\right)^{1/2}\right)  \\
&\qquad  \times \exp\left(- a \, \omega_1^2 + c  \, \omega_1 \omega_2 - b \, \omega_2^2  \right)   \,,
\end{split}
\end{align}
 where $\ell = R_0 e^{-\tau} \sqrt{\cosh(2 u \tau)}$. 
As in Sec. \ref{sec:resetBG}, Eqs. \eqref{pcircle} allow us to calculate the characteristic properties of the system for case $B$. To distinguish these quantities from their counterparts in case $A$, we introduce the superscript $R_0$, indicating that resetting occurs to a randomly selected point on the circumference of a circle of radius $R_0$.

\subsection{Moments and cross-moments of the position PDF}

Appropriately differentiating $\Phi_r^{\rm st}(\omega_1,\omega_2|R_0)$ in Eq.   \eqref{pcircle} with respect to variables $\omega_1$ and $\omega_2$,
and then setting $\omega_1 = \omega_2 = 0$, we have that the moments $\overline{x^n}^{(R_0)}$ and  $\overline{y^n}^{(R_0)}$ with odd $n$ vanish, as they should. Concurrently, the cross-moment $\overline{xy}^{(R_0)}$ does not vanish,
\begin{align}\label{eq:momentscircle}
\overline{xy}^{(R_0)} = - \frac{u \left(r R_0^2 + 2 (T_x + T_y) \right)}{(2 + r)^2 - 4 u^2} \,,
\end{align}
as well as the even moments of positions. In particular, we have
\begin{align}\label{eq:momentscircle2}
\begin{split}
\overline{x^2}^{(R_0)} &= \frac{(2 + r)^2 \left(r R_0^2 + 4 T_x\right) - 8 u^2 (T_x - T_y)}{2 (2+r) \left((2+r)^2 - 4 u^2\right)} \,, \\
\overline{y^2}^{(R_0)}  &= \frac{(2 + r)^2 \left(r R_0^2 + 4 T_y\right) + 8 u^2 (T_x - T_y)}{2 (2+r) \left((2+r)^2 - 4 u^2\right)} \,,
\end{split}
\end{align}
such that the averaged moment of inertia can be easily inferred and is given by
\begin{align}
	\label{cross}
\overline{I}^{(R_0)} = \frac{(2 + r) \left(r R_0^2 + 2 (T_x + T_y)\right)}{(2+r)^2 - 4 u^2} \,.
\end{align}
It is instructive to compare Eqs. \eqref{eq:momentscircle2} and \eqref{cross} with the corresponding expressions in case $A$, Eqs. \eqref{eq:mom1}, \eqref{eq:mom2} and \eqref{eq:I}. The denominators of the respective moments appear to be exactly the same for both resetting protocols, while the numerators differ. This is not accidental: the denominators are determined solely by the relaxation dynamics of the underlying BG and the resetting rate $r$, and are therefore independent of the spatial distribution of the resetting positions. In contrast, the numerators encode the geometry of the resetting protocol through the moments of the resetting coordinates. Averaging uniformly over the circumference of a circle thus modifies the numerators without affecting the denominators. This also implies that the conditions for the existence of finite second moments are identical for the two resetting protocols.
\begin{figure}[h]
	\includegraphics[width=14cm]{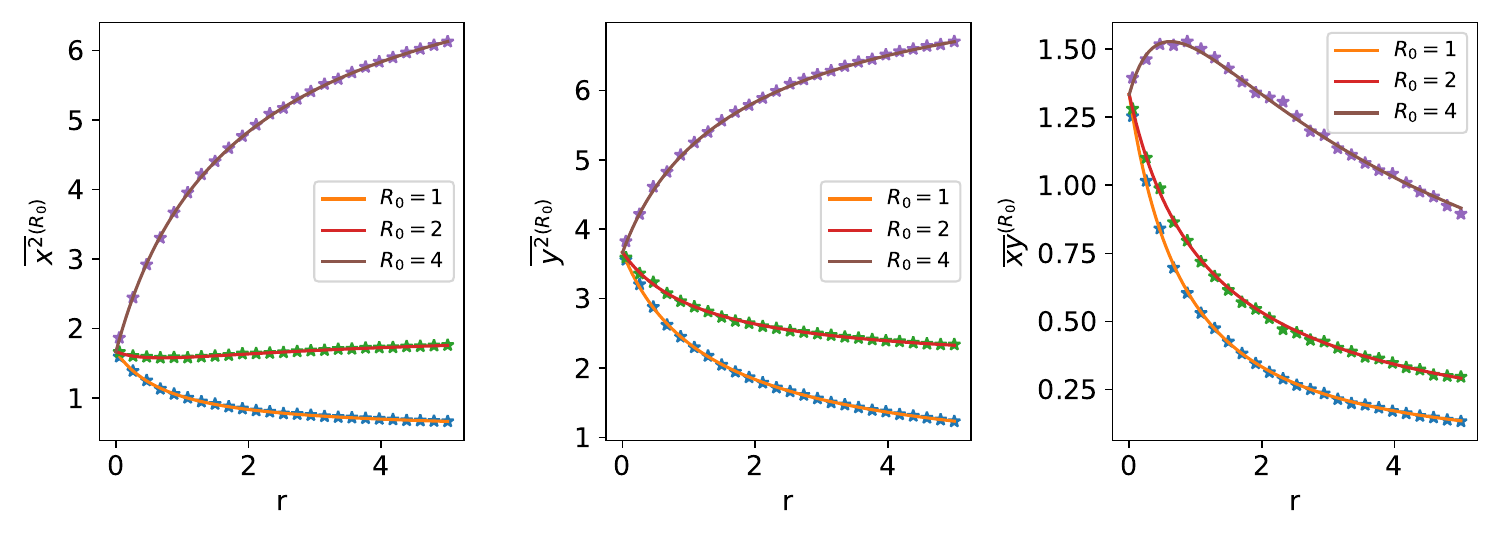}
	\caption{Case B. Second moments   $\overline{x^2}^{(R_0)}$,  $\overline{y^2}^{(R_0)}$ and the cross-moment $\overline{xy}^{(R_0)}$ versus the resetting rate $r$. The parameters $T_x=1$, $T_y=3$, $u=0.5$, and $R_0=1,2$ and $4$ (see the inset). Solid curves are theoretical predictions in Eqs. \eqref{eq:momentscircle} and  \eqref{eq:momentscircle2}, while the symbols depict the results of numerical simulations. }\label{fig:momentscircle}
\end{figure}

Figure~\ref{fig:momentscircle} compares the exact cross-moment in Eq.~\eqref{eq:momentscircle} and the second moments in Eq. \eqref{eq:momentscircle2} with simulations.  One observes that the results obtained in the case B
 differ qualitatively from those obtained for resetting to a fixed point. Indeed, 
the angular averaging eliminates the first moments and removes the directional
resetting contribution, since
$\langle X_0^2\rangle_{\rm circle}=\langle Y_0^2\rangle_{\rm circle}=R_0^2/2$
and $\langle X_0Y_0\rangle_{\rm circle}=0$.  Nevertheless, the coupling
between the two coordinates survives the angular average:
$\overline{xy}^{(R_0)}$ remains nonzero and has the sign opposite to $u$.
For $r\to\infty$,
$\overline{x^2}^{(R_0)},\overline{y^2}^{(R_0)}\to R_0^2/2$ and
$\overline{xy}^{(R_0)}\to0$.  Thus frequent resetting to a random  point on a circle restores the
isotropic second moments of the resetting ensemble, whereas frequent
fixed-point resetting retains the anisotropy of the selected resetting point.

A noteworthy feature of Eqs.~\eqref{eq:momentscircle} and
\eqref{eq:momentscircle2} is that their dependence on $r$ may again be
non-monotonic.  This is particularly transparent for the cross-moment.  Observe that
\begin{equation}\label{eq:dxycircle0}
 \left.\frac{d\overline{xy}^{(R_0)}}{dr}\right|_{r=0}
 =
 \frac{u\left[R_0^2u^2-R_0^2+2(T_x+T_y)\right]}
 {4(1-u^2)^2}.
\end{equation}
Hence, for $u>0$, the magnitude $|\overline{xy}^{(R_0)}|$ initially
\emph{increases} with $r$ when
\begin{equation}\label{eq:nonmonocircle}
 R_0^2(1-u^2)>2(T_x+T_y).
\end{equation}
Since $\overline{xy}^{(R_0)}\simeq-uR_0^2/r$ for $r\to\infty$, it must then
pass through an extremum at an intermediate resetting rate (which corresponds to the case $R_0=4$ depicted in the right panel of  Fig.~\ref{fig:momentscircle}).  The variances can
display the same type of non-monotonic crossover, depending on $R_0$ and on
the temperature anisotropy.  Physically, two effects compete.  At small and
moderate $r$, resetting repeatedly relocates the BG at a finite distance
$R_0$ from the origin and can therefore enhance positional fluctuations or
correlations.  At larger $r$, there is insufficient time for the coupled
Ornstein--Uhlenbeck dynamics to develop these correlations before the next
resetting event, and the moments approach those of the isotropic resetting ensemble.
This mechanism is closely related to the non-monotonic behavior found in
case A for resetting to a point: in both cases it results from competition
between relaxation in the confining potential and repeated reset.
The crucial difference is geometrical.  A fixed resetting point carries a
persistent directional memory through $X_0$, $Y_0$ and $X_0 Y_0$, whereas
the circular averaging removes this directional memory at every resetting event.  Thus
the non-monotonicity in case B is generated by the resetting \emph{radius} and
the subsequent anisotropic dynamics, rather than by the orientation of a
particular resetting point.

Having characterised the low-order moments, we next examine the full
stationary position distribution.  As in case A, this reveals spatial
features that cannot be inferred from the moments alone; in the circular
protocol these features are essentially non-Gaussian because the resetting
ensemble is itself extended.

\subsection{Position PDF for resetting to a circle}

The expression in Eq. \eqref{pcircle} for the position PDF
$P_r^{\rm st}(x,y|R_0)$ in case of a stochastic resetting to a random point on the circumference of a circle is too complicated to be evaluated exactly. To simplify the formula and make it amenable to a numerical analysis we use the following integral identity, which generalises the one in Eq.  \eqref{represent},
\begin{align}
	J_{0}\left(m \sqrt{\omega_1^2 + \omega_2^2 - 2 \cos(\phi) \omega_1 \omega_2}\right) = \frac{1}{2 \pi} \int^{\pi}_{-\pi} d\theta \, e^{i m \omega_1 \cos(\theta) - i m \omega_2 \cos(\theta - \phi)} \,.
	\end{align}
Inserting this identity into Eq.~\eqref{pcircle} and integrating over
$\omega_1$ and $\omega_2$, we obtain
\begin{align}
	\label{prob}
	P_r^{\rm st}(x,y|R_0) = \frac{r}{(2 \pi)^2} \int^{\infty}_0
	\frac{d\tau}{\sqrt{d}} e^{- r \tau}
	\int^{\pi}_{-\pi} d\theta
	\exp\left(- \frac{b D_x^2 + c D_x D_y + a D_y^2}{d}\right) \,.
\end{align}
with
\begin{align}
	\begin{split}
		D_x = x + \ell \cos(\theta) \,, \quad
		D_y  = y - \ell \cos(\theta - \phi) \,,\quad
		\phi = \arccos\left(\tanh(2 u \tau)\right) \,,
		\end{split}
	\end{align}
	where $\ell$ is defined below Eq. \eqref{pcircle}.

\begin{figure}[h]
	\centering
	\includegraphics[width=50mm]{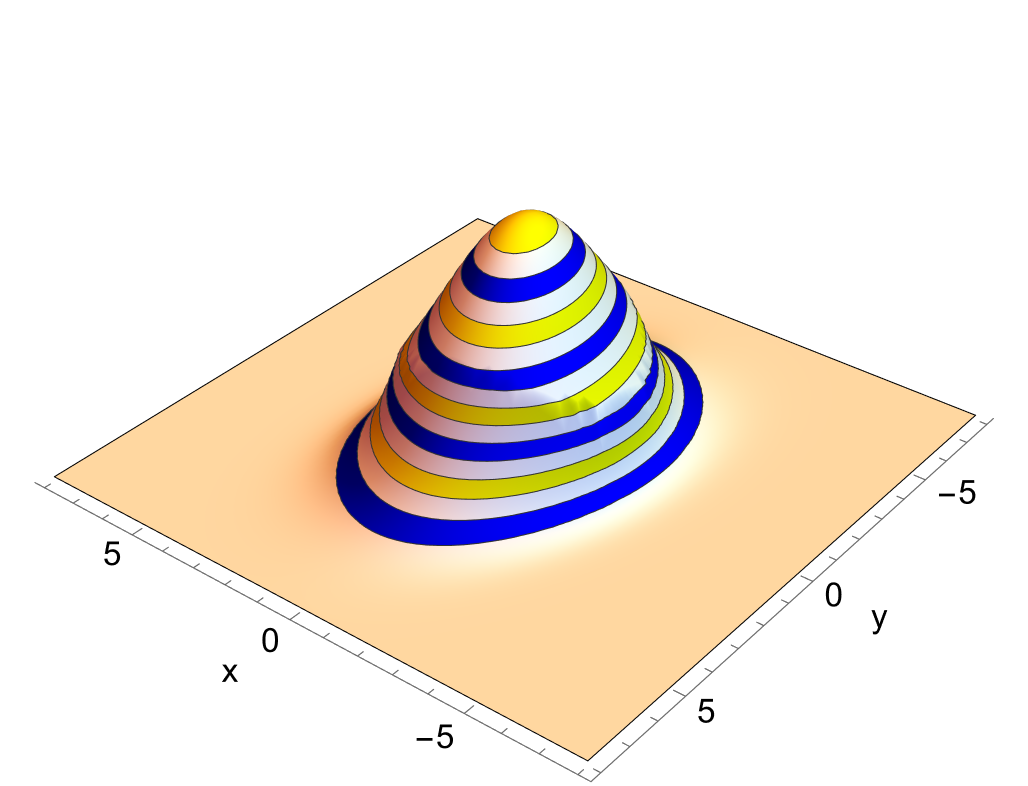}  
	\includegraphics[width=50mm]{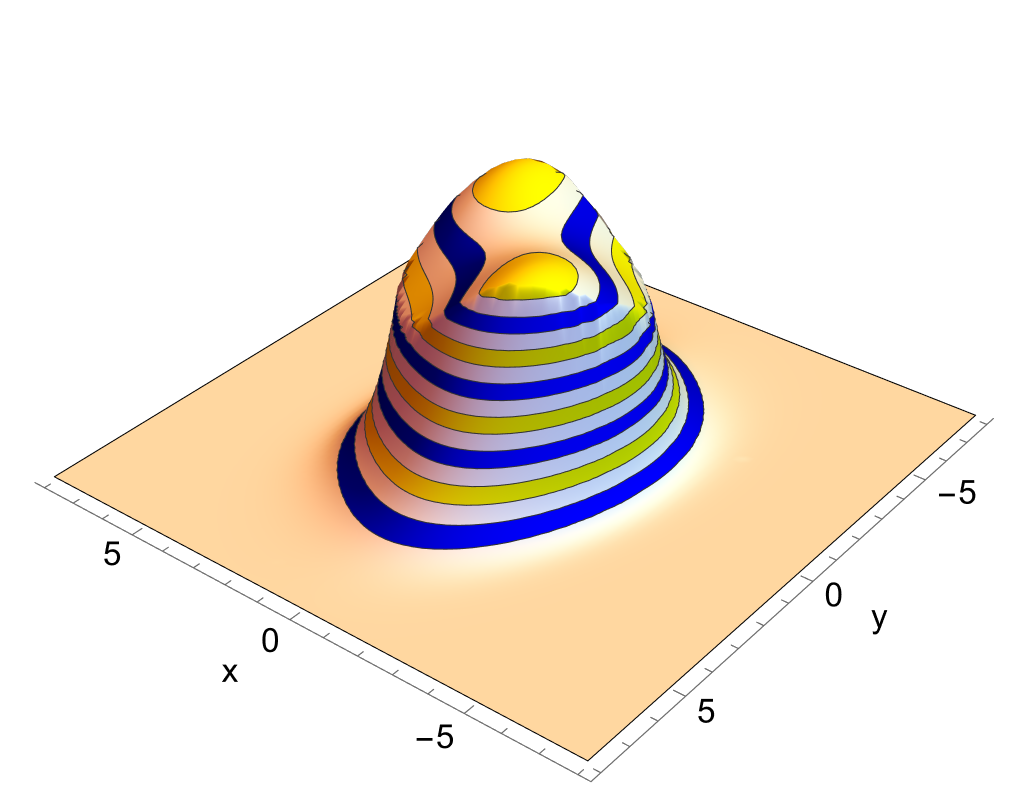}
	\includegraphics[width=50mm]{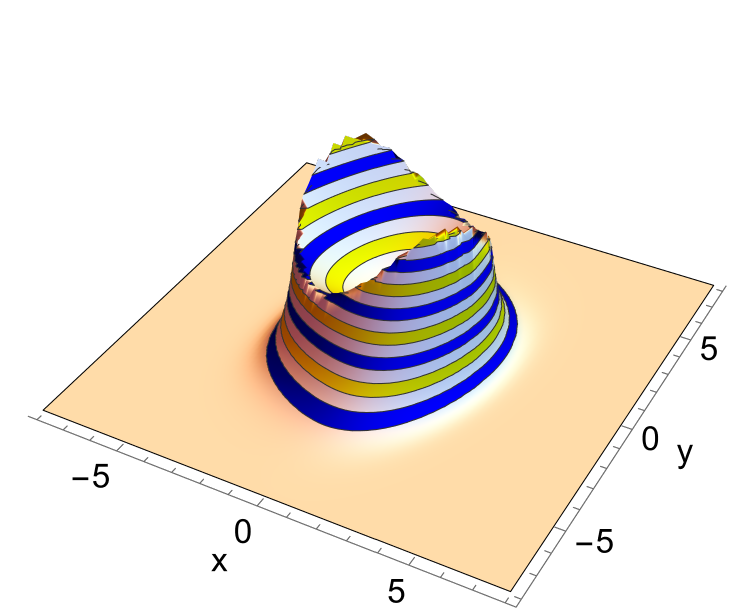}
	\includegraphics[width=50mm]{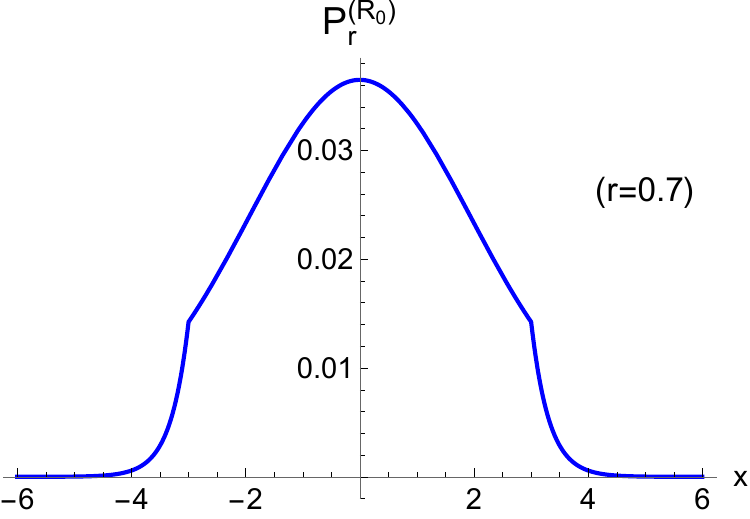}\includegraphics[width=50mm]{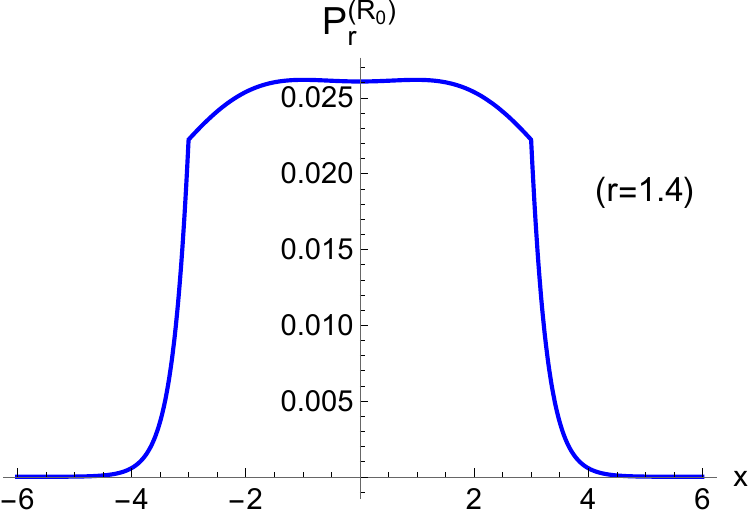} 	\includegraphics[width=50mm]{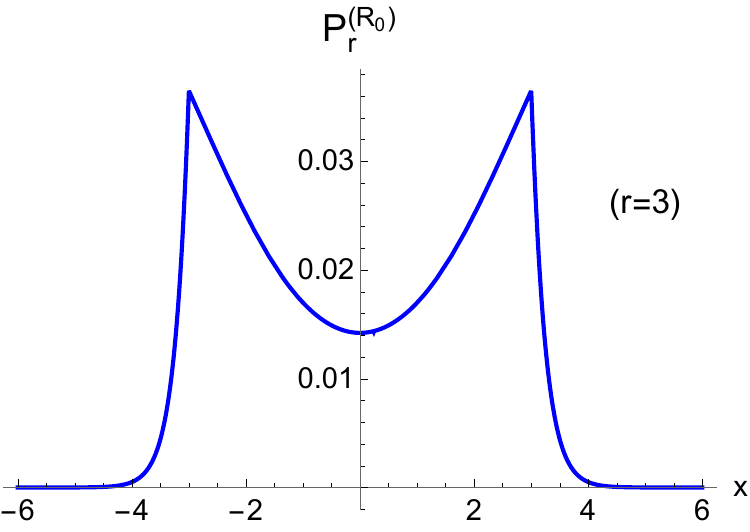} 
	\caption{Case B. The position PDF for $R_0 = 3$, $T_x = 1$, $T_y = 4$ and $u = 0.5$. Top row: $P^{\rm st}_r(x,y|R_0)$ in Eq. \eqref{prob}  for  $r = 0.7$ (left panel), $r = 1.4$ (middle panel) and $r = 3$ (right panel). Bottom row: The cut of the joint PDF at $y=0$:  $P^{(R_0)}_r = P^{\rm st}_r(x,y=0|R_0)$ as a function of $x$ for the same values of $r$ (see the inset). }
	\label{fig:11} 
\end{figure}

Equation~\eqref{prob} can be evaluated numerically.  Figure~\ref{fig:11}
shows the progressive reorganisation of the stationary density for the case of resetting to a randomly chosen point on a circle.  At
$r=0.7$ the three-dimensional surface still retains a visible memory of the
elliptic BG density, but the cut at $y=0$ already displays two
symmetric non-analytic features close to the resetting circle.  This is in sharp
contrast with case A, where a single off-center resetting point produces one
distinguished cusp and an explicitly displaced stationary state.
At $r=1.4$ the competition between confinement and resetting is strongest:
the central region becomes relatively flat and saddle-like, while the
finite-distance structures associated with the resetting circle remain visible.
At $r=3$ the center is depleted and probability accumulates near the resetting
circle.  This is the natural large-$r$ regime: the age since the last resetting event
is typically of order $r^{-1}$, so most trajectories have not had enough
time to move far away from the circumference on which they were injected.

The stationary position PDF thus exhibits a nontrivial spatial structure resulting from the interplay between circular resetting and the anisotropic dynamics of the BG. Although the resetting positions are uniformly distributed along a circle, the subsequent evolution is governed by the coupling between the coordinates and by the unequal temperatures $T_x$ and $T_y$. Consequently, the probability density does not retain the circular symmetry of the resetting protocol, nor the elliptic symmetry characteristic of the BG without resetting. Instead, its shape depends on the relative importance of resetting and relaxation, controlled by the resetting rate $r$.

An important difference with case A is that circular resetting does not introduce a preferred direction or displace the mean position away from the origin. Nevertheless, it substantially modifies the stationary distribution, particularly at intermediate resetting rates, where the contributions from trajectories that have undergone different periods of relaxation coexist. As $r$ increases, the influence of the resetting geometry becomes progressively more pronounced, ultimately leading to the accumulation of probability near the resetting circle, as observed in Fig. \ref{fig:11}.

\subsubsection{Marginal position PDFs.}

The marginal position PDFs can be obtained directly from the
full two-dimensional PDF in Eq.~\eqref{pcircle}. 
For the $x$ and $y$ marginals, setting $\omega_2=0$ or $\omega_1=0$ in Eq.~\eqref{pcircle} gives
\begin{align}\label{eq:pxcircleFourier}
	\begin{split}
 P_{r}^{\rm st}(x|R_0)
& =\frac{r}{\pi}\int_0^\infty d\tau\,e^{-r\tau}
 \int_{0}^{\infty} d\omega\,
 e^{-a \omega^2}
 J_0\!\left(\ell \omega\right) \,  \cos(\omega x) , \\
  P_{r}^{\rm st}(y|R_0)
& =\frac{r}{\pi}\int_0^\infty d\tau\,e^{-r\tau}
 \int_{0}^{\infty}d\omega\,
 e^{-b\omega^2}
 J_0\!\left(\ell \omega\right) \, \cos(\omega y)\,,
 \end{split}
\end{align}
where $\ell =R_0e^{-\tau}\sqrt{\cosh(2u\tau)}$.  For numerical purposes, it is convenient to replace the integration over the infinite range of $\omega$ by an integral over a finite interval. Using the standard angular representation of the Bessel function $J_0(z)$ \cite{AbramowitzStegun1964}, one finds
\begin{align}\label{eq:margcircleReal}
	\begin{split}
 P_{r}^{\rm st}(x|R_0)
 &=\frac{r}{2\pi}\int_0^\infty d\tau\,e^{-r\tau}
 \int_0^{2\pi}\frac{d\varphi}{\sqrt{4\pi a}}\,
 \exp\!\left[-\frac{(x-\ell \cos\varphi)^2}{4a}\right],
 \\
 P_{r}^{\rm st}(y|R_0)
 &=\frac{r}{2\pi}\int_0^\infty d\tau\,e^{-r\tau}
 \int_0^{2\pi}\frac{d\varphi}{\sqrt{4\pi b}}\,
 \exp\!\left[-\frac{(y-\ell \cos\varphi)^2}{4b}\right],
 \end{split}
\end{align}
where $a$ and $b$ are defined in Eq.~\eqref{coeft}. In fact, the remaining angular integral can also be evaluated analytically, yielding an infinite series of Bessel functions (see Appendix \ref{C}). This series can be truncated at a finite order to achieve any prescribed accuracy. For our numerical analysis, however, we find it more convenient to work directly with the exact integral representation in Eq. \eqref{eq:margcircleReal}.

\begin{figure}[h]
	\includegraphics[width=14cm]{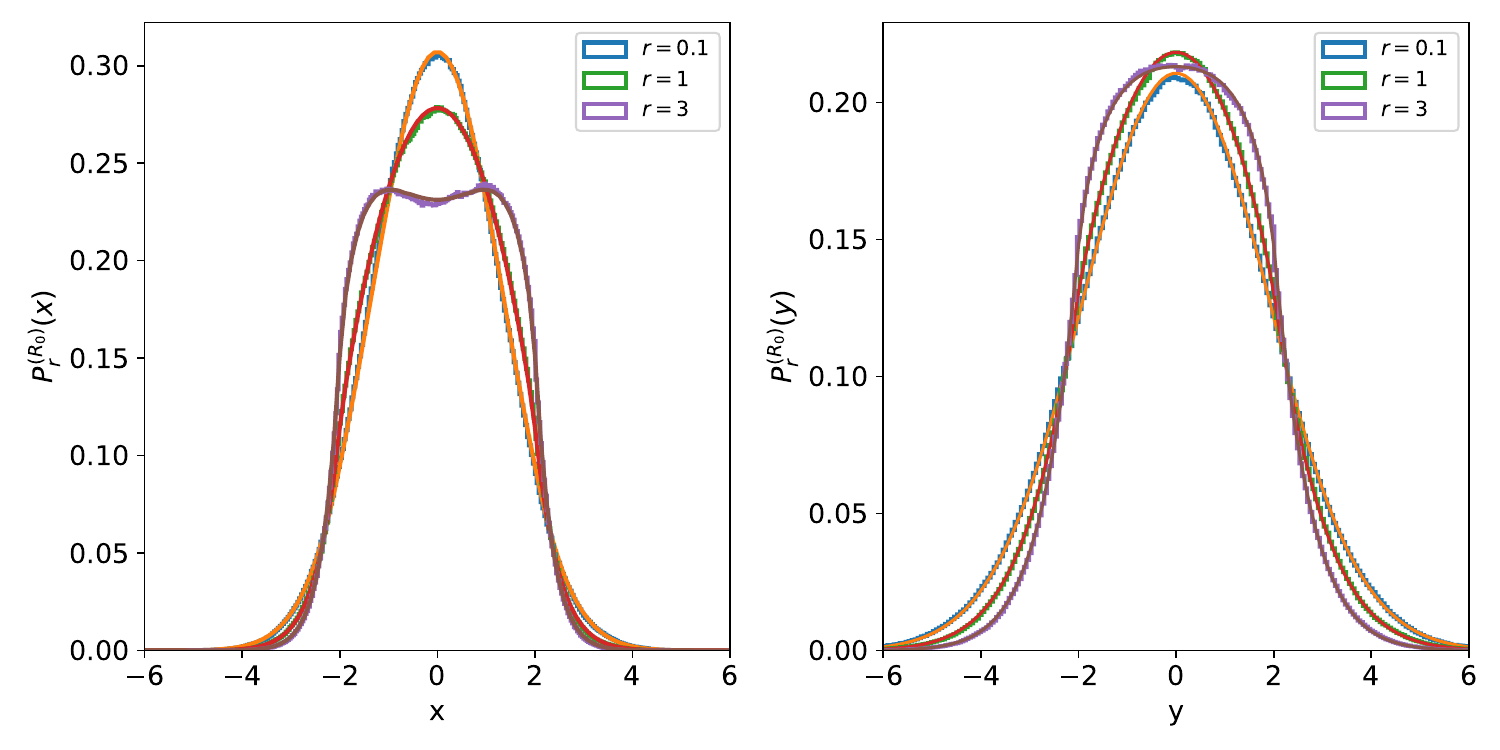}
	\includegraphics[width=14cm]{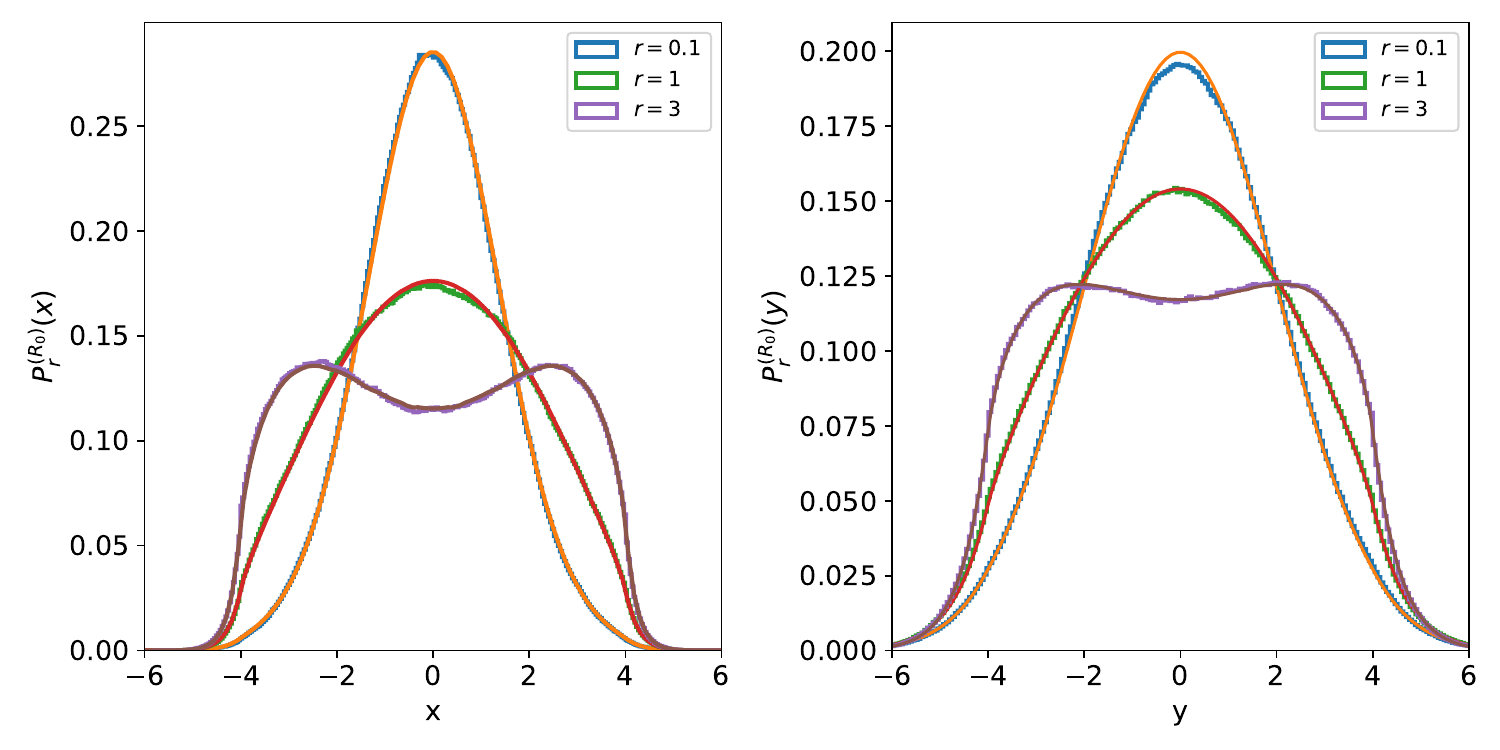}
	\caption{Case B. The marginal PDFs $P_{r}^{(R_0)}(x)=P_{r}^{\rm st}(x|R_0)$ (left column) and  $ P_{r}^{(R_0)}(y)=P_{r}^{\rm st}(y|R_0)$ (right column) for  $T_x=1$, $T_y=3$ and $u=0.5$. $R_0=2$ (first row) and $R_0=4$ (second row). The noisy magenta curves are the results of numerical simulations, while the  solid curves are analytical predictions in Eqs. \eqref{eq:margcircleReal} with numerically evaluated integrals.}\label{fig:Pmarginal}
\end{figure}

Figure \ref{fig:Pmarginal} presents the marginal PDFs $P_{r}^{\rm st}(x|R_0)$ (left column) and $P_{r}^{\rm st}(y|R_0)$ (right column) for $T_x=1$, $T_y=3$ and $u=0.5$. The solid curves represent our analytical predictions, obtained by numerically evaluating the two-fold integrals in Eqs.~\eqref{eq:margcircleReal}, while the fluctuating magenta curves show the results of numerical simulations. The upper and lower rows correspond to resetting radii $R_0=2$ and $R_0=4$, respectively.

For $R_0=2$, the marginal PDF $P_{r}^{\rm st}(x|R_0)$ undergoes a pronounced change in shape as the resetting rate increases. At $r=0.1$, the distribution is approximately Gaussian, whereas at $r=3$ it develops a bimodal structure, with two symmetric maxima separated by a minimum at the origin. In contrast, $P_{r}^{\rm st}(y|R_0)$ remains unimodal over the same range of resetting rates, retaining its maximum at the origin.
For the larger resetting radius $R_0=4$, these effects become more pronounced. The $x$-marginal exhibits the same qualitative evolution, from an approximately Gaussian distribution at small $r$ to a distinctly bimodal one at $r=3$, with two well-defined symmetric maxima at nonzero values of $x$. Interestingly, the positions of these maxima do not coincide with $\pm R_0$, indicating that the most probable values of $x$ are not determined solely by the geometry of the resetting circle, but also by the subsequent relaxation dynamics. The dependence of the $y$-marginal on the resetting rate becomes more pronounced as well, illustrating the different responses of the two coordinates to circular resetting when $T_x \neq T_y$.

\subsection{Probability current and vorticity}

As in case A, the local probability current accounts only for the continuous motion of the particle between successive resetting events. The instantaneous jumps to the resetting circle are not included in this current.

\begin{figure}[h]
	\centering
	\includegraphics[width=50mm]{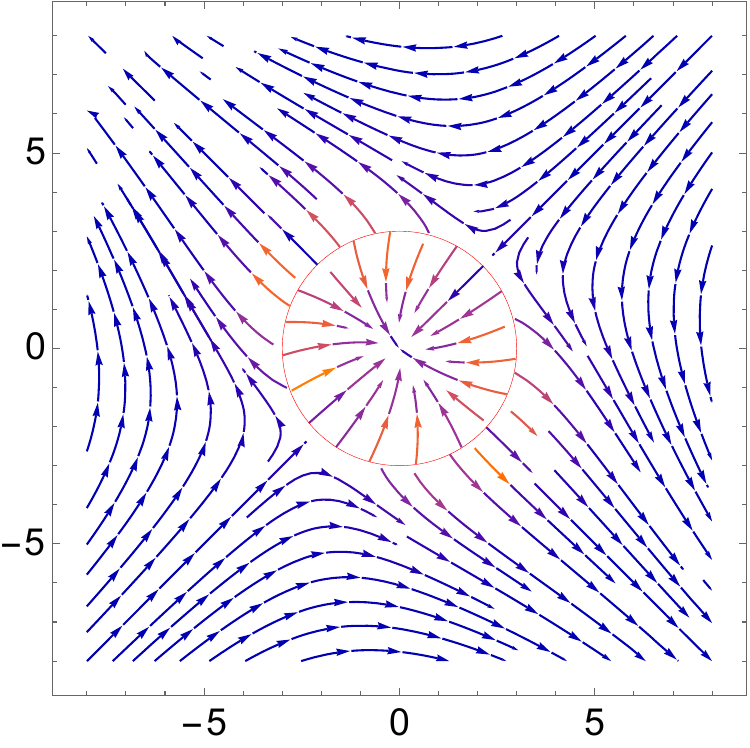}  \includegraphics[width=50mm]{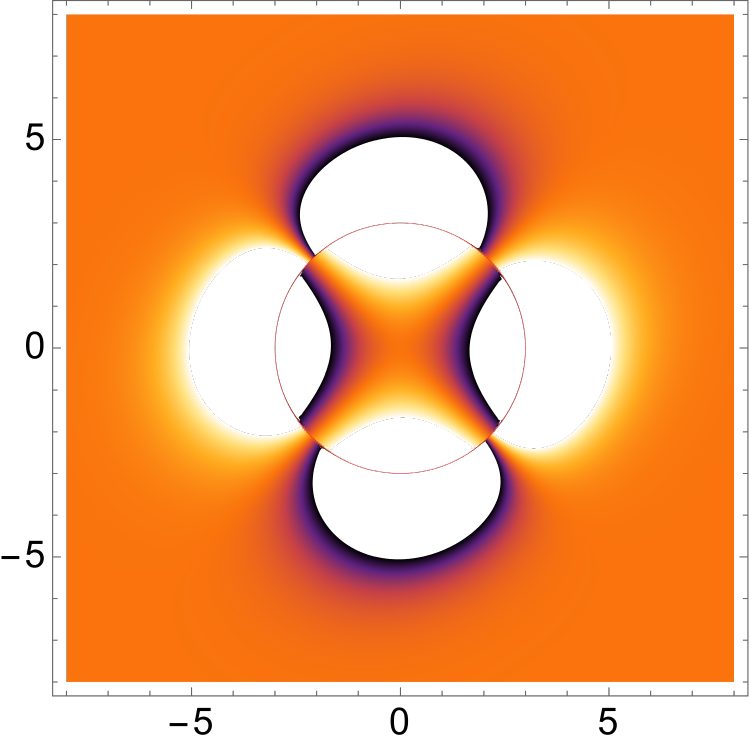}
	\includegraphics[width=14mm]{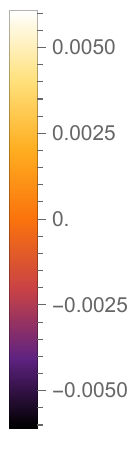}
	\caption{Case B. 
		Left panel: The stream plot of the probability current $\vec{j}^{(R_0)}_r$ on the $(x,y)$-plane for the resetting rate $r = 3$. Right panel: The density plot of the vorticity of $\vec{j}^{(R_0)}$ for $r = 3$. The parameters: radius of the circle (indicated by a thin red curve) $R_0 = 3$, $u = 0.5$ and $T_x = T_y = 4$ (suppressed rotation).}
	\label{fig:22} 
\end{figure}

In Fig.~\ref{fig:22}, we isolate the contribution of circular resetting to the stationary probability current by setting $T_x=T_y$. Recall that, for equal temperatures, the BG without resetting satisfies detailed balance and its stationary probability current vanishes. Therefore, the streamlines and vorticity shown in this figure arise entirely from stochastic resetting. This makes the comparison with Fig.~\ref{fig:C}
of case A particularly instructive.  A fixed resetting point acts as a localized
source and creates an off-center flow pattern.  Here the source is spread
uniformly over a circle and no point on its circumference
is privileged.
Nevertheless, the resulting flow does not exhibit circular symmetry. Due to the coupling $u$ between the two coordinates, the particle does not simply move radially towards the origin after each resetting event. Instead, its subsequent motion depends on its position on the resetting circle. Although resetting occurs uniformly along the circle, the resulting probability current retains a nontrivial spatial structure, giving rise to the lobed patterns of streamlines and vorticity observed in Fig.~\ref{fig:22}. These patterns reflect the combined effects of the circular resetting protocol and the coupled relaxation dynamics.
The most pronounced variations of the probability current and vorticity occur in the vicinity of the resetting circle, $\rho=R_0$. Although this circle does not represent a physical boundary and the particle can cross it freely, it plays a special role in the dynamics as the set of possible resetting positions. At large resetting rates, the particle spends most of its time close to the circle, since the intervals between successive resetting events are typically short. This leads to strong spatial variations of the local probability current and to the appearance of neighboring regions with positive and negative vorticity. These features reflect the redistribution of probability by the continuous dynamics between resetting events.

\begin{figure}[h]
	\centering
	\includegraphics[width=45mm]{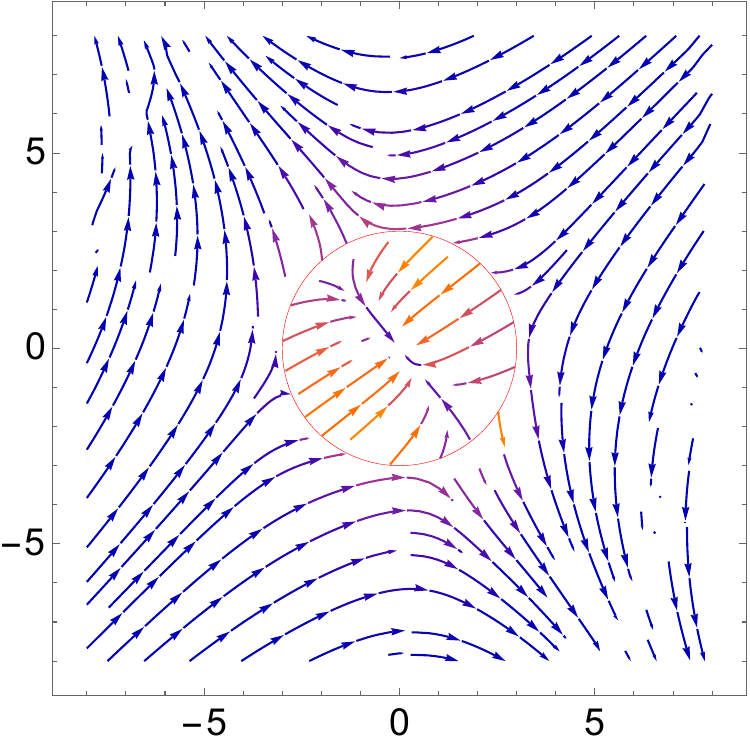} 
	\includegraphics[width=45mm]{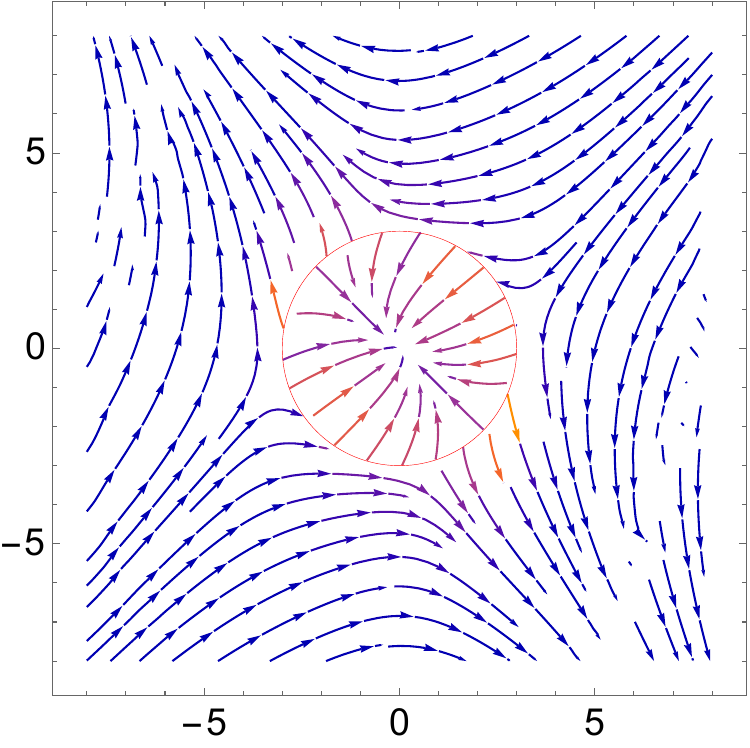}
	\includegraphics[width=45mm]{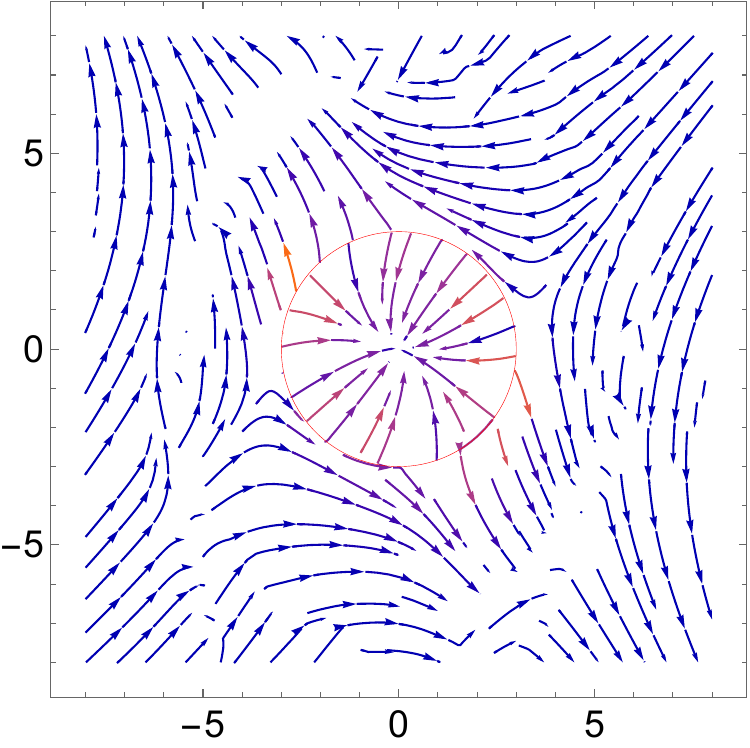}
	\caption{Case B. 
		The stream plot of the probability current $\vec{j}^{(R_0)}_r$ on the $(x,y)$-plane for the resetting rate $r = 0.7$ (left panel), $r = 1.4$ (middle panel) and $r = 3$ (right panel). The parameters: radius of the circle (indicated by a thin red curve) $R_0 = 3$, $T_x = 1$, $T_y = 4$ and $u = 0.5$.}
	\label{fig:23} 
\end{figure}

Figure~\ref{fig:23} shows the stationary probability current and its vorticity for unequal temperatures, $T_x\ne T_y$, when both the intrinsic BG dynamics and circular resetting contribute to the flow. At the relatively small resetting rate $r=0.7$, the current retains some features of the rotational motion characteristic of the BG without resetting. As $r$ increases, however, the influence of resetting becomes more pronounced, leading to substantial changes in both the streamlines and the spatial distribution of vorticity. At $r=3$, the flow exhibits a more complicated structure, with distinct patterns appearing inside and outside the resetting circle. The asymmetry of these patterns reflects the combined effects of the coupling $u$ and the temperature difference $\Delta_T= T_x-T_y$.

It is worth emphasizing that, unlike resetting to a fixed point considered in case A, circular resetting does not select a preferred spatial direction. Nevertheless, the resulting probability current is not rotationally symmetric, since the dynamics between successive resetting events depends on the coupling and the temperatures along the two coordinate axes. The nontrivial flow patterns observed in Figs.~\ref{fig:22} and \ref{fig:23} thus demonstrate how circular resetting modifies the spatial structure of the BG probability current, both in the absence and in the presence of a temperature difference.

\subsection{Angular momentum for resetting to a circle}

Averaging the fixed-point result in Eq.~\eqref{eq:Lmean} over
$(X_0,Y_0)=(R_0\cos\varphi_0,R_0\sin\varphi_0)$ gives
\begin{equation}\label{eq:Lcircle}
\overline{\cal L}^{(R_0)}
 =\frac{2u}{2+r}(T_y-T_x).
\end{equation}
The resetting-position contribution cancels because
$\langle Y_0^2-X_0^2\rangle_{\rm circle}=0$.  Hence the result is
independent of $R_0$, unlike the fixed-point case where resetting may even
reverse the sign of the mean rotation.  Circular resetting only attenuates
the thermal BG contribution, from $u(T_y-T_x)$ at $r=0$ to
zero for $r\to\infty$.

\begin{figure}[h]
	\centering
	\includegraphics[width=7cm]{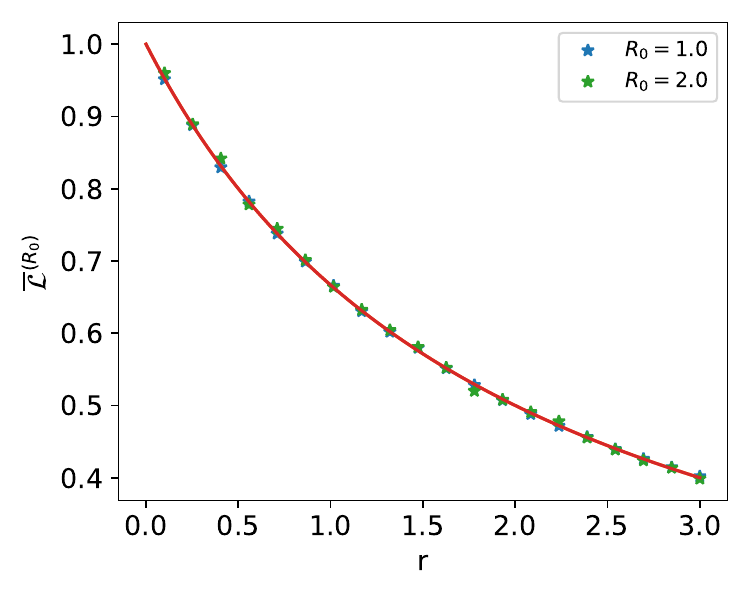}
	\includegraphics[width=7cm]{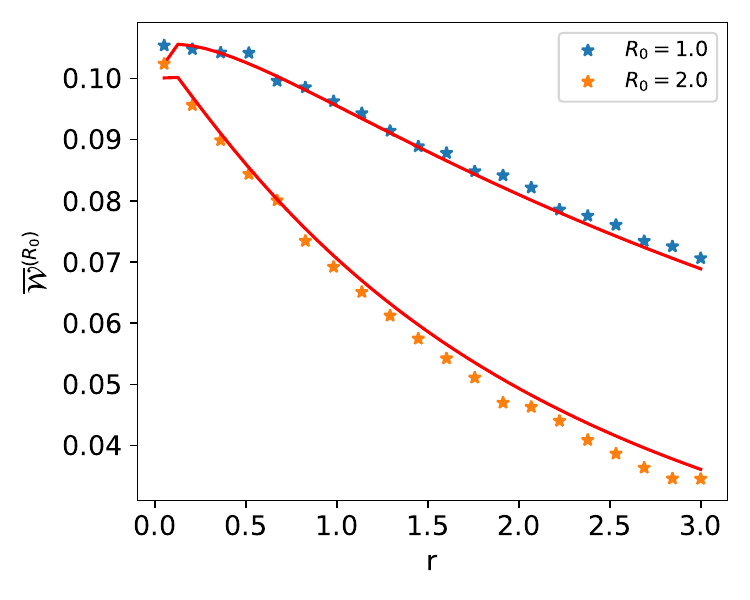}
	\caption{Case B. Left panel: mean angular momentum
		$\overline{\cal L}^{(R_0)}$ versus $r$ for $T_x=1$, $T_y=3$, $u=0.5$ and
		$R_0=1,2$. The solid curve is Eq.~\eqref{eq:Lcircle}, symbols are
		simulations. Right panel: mean angular velocity $\overline{\cal W}^{(R_0)}$ for
		the same parameters. Unlike the angular momentum, the angular velocity
		retains a dependence on $R_0$.}\label{fig:Lcircle}
\end{figure}

The left panel of Fig.~\ref{fig:Lcircle} provides a direct test of this symmetry argument: the simulation points for $R_0=1$ and $R_0=2$ lie on the same curve despite the different resetting radii. The mean decreases in magnitude with $r$ and retains the sign fixed by $u(T_y-T_x)$. Thus the radial scale of resetting does not affect $\overline{\cal L}$.  

The
full PDF of $\cal L$ is the angular average
\begin{equation}\label{eq:PLcircleaverage}
 P_r^{\rm st}({\cal L}|R_0)
 =\frac{1}{2\pi}\int_0^{2\pi}d\varphi_0\,
 P_r^{\rm st}({\cal L}|R_0\cos\varphi_0,R_0\sin\varphi_0).
\end{equation}

For finite observation step $\delta t$, its far tails are exponential, such that
\begin{equation}\label{eq:LcircleTail}
 \ln P_r^{\rm st}({\cal L}|R_0)
 \underset{{\cal L}\to\pm\infty}{\sim}
 -\lambda_\pm |{\cal L}|+o(|{\cal L}|),
\end{equation}
where the explicit expressions for the decay rates are derived in Appendix~\ref{appendix:momentum}. The leading exponential decay is independent of the resetting radius $R_0$ and the resetting rate $r$, and is determined entirely by the large-distance Gaussian behavior of the BG without resetting. The resetting parameters $R_0$ and $r$ affect only the subleading terms in the asymptotic expressions.

\begin{figure}[t]
	\includegraphics[width=15cm]{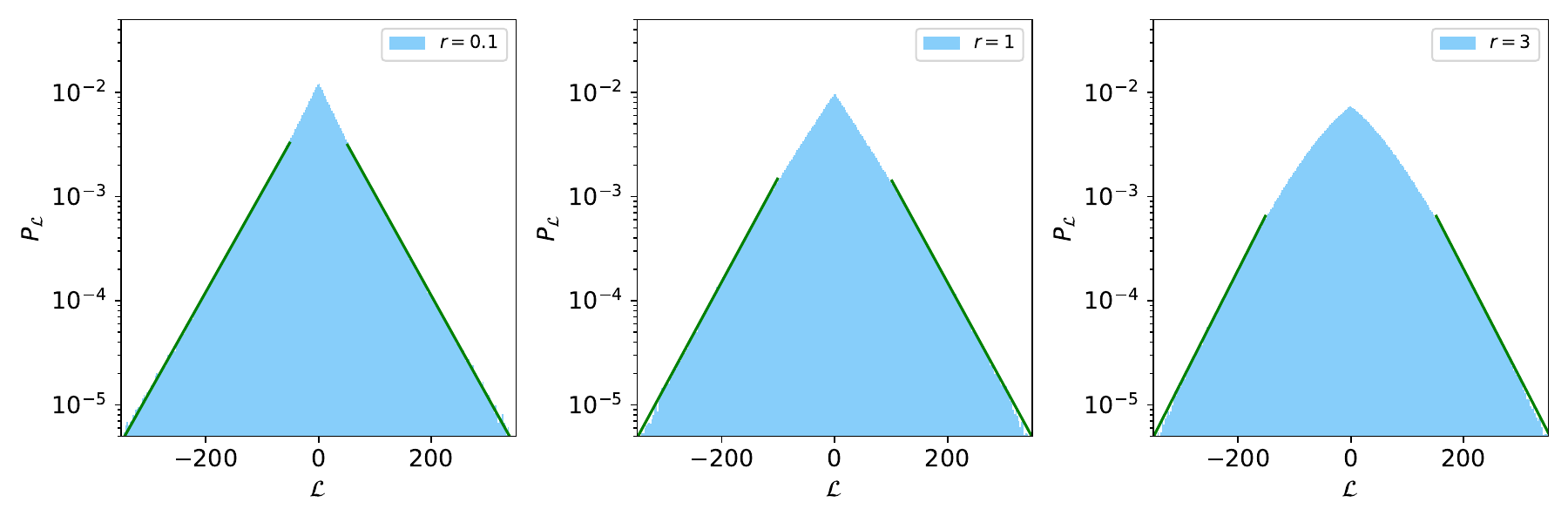}
	\caption{Case B: The PDF $P_{\cal L} = P_r^{\rm st}({\cal L}|R_0)$ of the specific angular momentum $\cal L$ for $T_x=1$, $T_y=3$ and $u=0.5$. The radius $R_0=3$ and the reset rates
		are $r=0.1$ (left panel), $r= 1$ (middle panel) and $r = 3$ (right panel). The histograms are the PDFs evaluated in numerical simulations. The green curves show the exponential tails of the PDF predicted by Eq.~\eqref{eq:LcircleTail}.}\label{fig:PdfLcircle}
\end{figure}

Figure \ref{fig:PdfLcircle} presents the stationary PDF of the specific angular momentum $\mathcal{L}$ for three resetting rates: $r=0.1$ (left panel), $1$ (middle panel), and $3$ (right panel), with the resetting positions uniformly distributed on a circle of radius $R_0=3$. For weak resetting, the distribution exhibits a nearly cusp-like maximum near $\mathcal{L}=0$, together with asymmetric tails extending over both positive and negative values of $\mathcal{L}$. As $r$ increases, the central peak becomes progressively broader and more rounded, while the overall distribution retains its asymmetry. Remarkably, the far tails remain exponential for all three resetting rates, as evidenced by their approximately linear behavior on the logarithmic scale. The red lines represent the leading asymptotic behavior predicted by Eq. \eqref{eq:LcircleTail}, showing good agreement with the numerical results. In particular, the slopes of these tails remain essentially unchanged as $r$ increases, consistent with the fact that their leading exponential decay rates are independent of both $r$ and $R_0$.

It is instructive to compare these results with those in Fig. \ref{fig:L}, corresponding to resetting to a fixed point. In both cases, the PDFs display asymmetric exponential tails, indicating that large fluctuations of angular momentum remain possible irrespective of the resetting protocol. The essential difference lies in the influence of the resetting geometry: For resetting to a fixed point, the preferred spatial direction introduces an additional contribution to the mean angular momentum, which may either reinforce or oppose the intrinsic gyration and can even reverse its sign as the resetting rate increases. In contrast, uniform resetting on a circle averages out this directional contribution, so that the mean angular momentum retains its original sign and decreases monotonically in magnitude with $r$, as shown in Fig. \ref{fig:Lcircle}. Thus, although circular resetting substantially reshapes the central part of the angular momentum distribution, it neither reverses the mean rotation nor alters the leading exponential decay of its far tails. This illustrates the distinct roles played by the resetting geometry in controlling typical rotational behavior and rare, large angular momentum fluctuations.	
	
\subsection{Angular velocity for resetting to a circle}

Angular averaging of the fixed-point expression gives
\begin{equation}\label{eq:WcircleMean}
 \overline{\cal W}^{(R_0)}
 =u\int_{-\infty}^{\infty} \int_{-\infty}^{\infty} dx\,dy\, \frac{y^2-x^2}{x^2+y^2} \, P_r^{\rm st}(x,y|R_0) \,.
\end{equation}
The explicit averaged integral is given in
Appendix ~\ref{appendix:Pdew}.  Unlike $\overline{\cal L}^{(R_0)}$, this
quantity cannot be reduced to finitely many position moments, and its
dependence on $R_0$ therefore survives angular averaging.

The dependence of the mean angular velocity $\overline{\cal W}$ on the resetting rate $r$ is shown in the right panel of Fig. \ref{fig:Lcircle} and contrasts with that of the mean angular momentum $\overline{\cal L}$ (left panel in this figure). In particular, unlike $\overline{\cal L}$, the curves for $\overline{\cal W}$ corresponding to different resetting radii do not collapse onto a single curve. This distinction can be understood by noting that, while angular averaging eliminates the directional contribution of resetting to the second moments entering $\overline{\cal L}$, the mean angular velocity involves an additional nonlinear spatial factor $1/(x^2+y^2)$, which preserves the dependence on the resetting radius. As $r$ increases, the particle spends an increasing fraction of time in the vicinity of the resetting circle, making the typical distance from the origin, and hence the value of $R_0$, increasingly relevant to $\overline{\cal W}$. Thus, the two panels highlight the different sensitivities of angular momentum and angular velocity to the geometry of resetting, demonstrating that these observables provide complementary characterizations of the stationary rotational dynamics.

For $R_0>0$ the stationary position density is regular at the origin, and
the same local calculation as for a fixed resetting point yields
\begin{equation}\label{eq:WcircleTail}
 P_r^{\rm st}({\cal W}|R_0)
 \underset{|{\cal W}|\to\infty}{\simeq}
 \frac{\pi(T_x+T_y)}{\delta t}\,
 \frac{P_r^{\rm st}(0,0|R_0)}{|{\cal W}|^3}.
\end{equation}
Thus, circular resetting does not change the asymptotic power-law decay $P_r^{\rm st}({\cal W}|R_0) \sim |W|^{-3}$. Remarkably, this is exactly the same power-law decay as for resetting to a fixed point (case A) and, more importantly, as for the BG without resetting \cite{Viot2024}. The resetting mechanism therefore affects only the prefactor of the power law, leaving its exponent unchanged. Consequently, the first moment of the angular velocity is finite, whereas the second and all higher-order moments diverge. This heavy-tailed distribution implies that the angular velocity exhibits highly intermittent fluctuations along individual trajectories, with occasional extremely large bursts of angular motion. Such bursts are associated with rare passages of the particle close to the origin, where even small displacements can produce very large changes in the angular position.

\begin{figure}[h]
 \includegraphics[width=15cm]{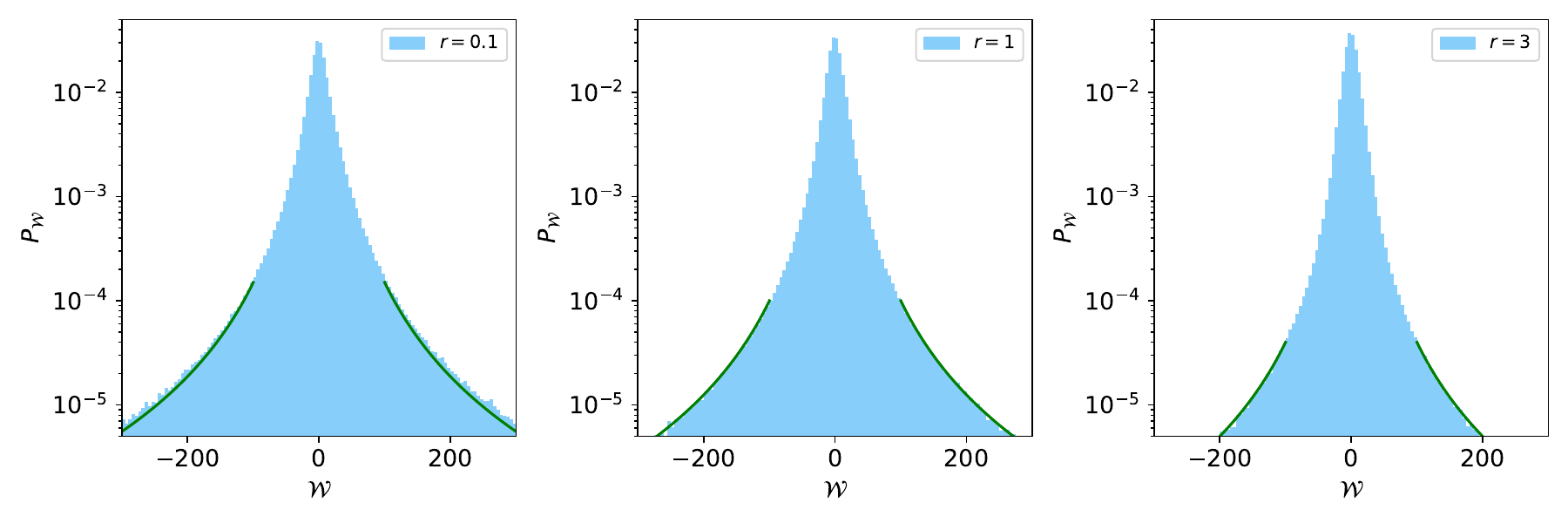}
	\caption{Case B: The PDF of the specific angular velocity $P_{\cal W} = P_r^{\rm st}({\cal W}|R_0)$  for $T_x=1$, $T_y=3$ and $u=0.5$. The reset parameters are $R_0 = 3$ and $r=0.1$ (left panel), $r=1$ (middle panel) and $r =3$ (right panel). The histograms present the PDFs obtained in numerical simulations.  The green curves show the power-law tails of the PDF predicted by Eq.~\eqref{eq:WcircleTail}.}\label{fig:PdfWcircle}
\end{figure}

Figure~\ref{fig:PdfWcircle} presents the numerically determined PDFs of the angular velocity ${\cal W}$ for  $R_0=3$  and three resetting rates: $r=0.1$, $1$, and $3$. As the resetting rate increases, the distributions become narrower. Nevertheless, all distributions exhibit the same asymptotic power-law decay, $P_r^{\mathrm{st}}({\cal W}|R_0)\sim |{\cal W}|^{-3}$, in agreement with Eq.~\eqref{eq:WcircleTail}. The green lines indicate the predicted asymptotic behavior. The amplitude of the tail depends on the resetting parameters, through the stationary position PDF evaluated at the origin, as shown explicitly in Eq.~\eqref{eq:WcircleTail}. Thus, circular resetting changes the probability of observing very large angular velocities but does not eliminate the heavy tails responsible for the divergence of the second and higher-order moments. 

\section{Conclusion}
\label{sec:conclusions}

In this work, we have studied how stochastic resetting modifies the stationary properties and rotational dynamics of the Brownian gyrator (BG). We considered two resetting protocols, differing in their spatial symmetry: resetting to a fixed point in the plane and resetting to a randomly selected point on a circle centered at the origin. By combining analytical calculations based on the renewal approach with numerical simulations, we have characterised the resulting stationary position distributions, their moments and correlations, local probability currents and vorticity, as well as the statistical properties of angular momentum and angular velocity. Our central finding is that resetting does not merely suppress or enhance the gyration: depending on the case at hand, it can qualitatively reorganise the stationary state and even reverse the mean rotational motion, while leaving certain properties of extreme fluctuations unchanged.

At the level of the stationary position distribution, resetting produces substantial departures from the Gaussian, elliptically symmetric state of the conventional BG. For resetting to a fixed point, probability accumulates around the resetting position, whereas uniform resetting on a circle can generate ring-shaped distributions with non-elliptic contours. The corresponding moments and correlations exhibit a rich dependence on the resetting rate, including non-monotonic behavior arising from the competition between thermal fluctuations and resetting-induced spatial excursions. In particular, resetting can generate correlations between the two coordinates even in the absence of direct dynamical coupling, illustrating how a common sequence of resetting events can correlate otherwise independent degrees of freedom. We have also shown that resetting generates non-vanishing local probability currents and vorticity even when the two temperatures are equal, so that the underlying BG dynamics would satisfy detailed balance. Resetting therefore acts as an independent source of non-equilibrium probability flows, whose spatial organization depends sensitively on the resetting protocol.

The distinction between the two resetting geometries becomes particularly striking when considering the rotational dynamics. Resetting to a fixed point introduces a preferred spatial direction, whose contribution to the mean angular momentum may compete with that generated by the temperature difference. For suitable parameters, this competition leads to a reversal of the mean angular momentum, demonstrating that resetting can change not only the magnitude but also the direction of the average gyration. This possibility disappears for uniform resetting on a circle: angular averaging eliminates the directional contribution, leaving a mean angular momentum that is independent of the resetting radius and decreases monotonically in magnitude with increasing resetting rate. The mean angular velocity, however, retains a dependence on the radius, reflecting its greater sensitivity to the spatial distribution of probability near the origin. Thus, the two angular observables respond differently to the geometry of resetting and capture complementary aspects of the stationary rotational dynamics.

An equally important conclusion concerns the fluctuations of these observables. For both resetting protocols, the angular-momentum PDFs exhibit exponential tails, whereas the angular-velocity PDFs retain the inverse-cubic algebraic decay characteristic of the BG without resetting. Although resetting modifies the shapes of these distributions and the amplitudes of their tails, it does not change their leading decay exponents. In particular, the persistence of the inverse-cubic tails implies that the second and higher-order moments of the angular velocity remain divergent. Hence, even frequent resetting cannot suppress the exceptionally large angular-velocity fluctuations associated with passages of the particle close to the origin. This provides a clear distinction between the effects of resetting on typical rotational motion, which can be strongly modified and even reversed, and on extreme angular-velocity fluctuations, whose asymptotic behavior remains robust.

Taken together, our results demonstrate that the spatial geometry of resetting provides an effective means of controlling the stationary structure and mean rotational response of the BG, but does not regularize its anomalously large angular-velocity fluctuations. This distinction between the controllability of average observables and the robustness of their extreme fluctuations is likely to be relevant beyond the particular model considered here. Natural extensions include resetting to more general spatial distributions, in particular nonuniform distributions along a circle, and the analysis of the thermodynamic costs associated with different resetting protocols. Another interesting direction concerns the role of finite temporal resolution in experimentally accessible measurements of rotational fluctuations.

\section*{References}

\providecommand{\newblock}{}

\appendix
\renewcommand{\thesection}{\Alph{section}}
\section{Probability density function of the specific angular momentum}
\label{appendix:momentum}

\subsection*{Finite-time-step definition and conditional distribution}

The Langevin trajectories are not differentiable, so that the angular momentum ${\cal L}$ in Eq.~\eqref{L} has to be appropriately redefined 
using a finite forward time step $\delta t$ \cite{Viot2024}. Between resetting events, the forward discretisation of Eq.~\eqref{Lan} reads
\begin{equation}\label{eq:discXY}
\begin{split}
 \frac{x_{t+\delta t}-x_t}{\delta t}
 &=-x_t-u y_t+\sqrt{\frac{2T_x}{\delta t}}\,\eta_x(t),\\
 \frac{y_{t+\delta t}-y_t}{\delta t}
 &=-y_t -u x_t+\sqrt{\frac{2T_y}{\delta t}}\,\eta_y(t),
\end{split}
\end{equation}
where $\eta_x$ and $\eta_y$ are independent standard Gaussian variables. The increment in Eq.~\eqref{eq:discXY} describes continuous BG motion during the observation step; the instantaneous displacement at a resetting event is not included. We represent then the angular momentum as
\begin{equation}\label{eq:Ldis}
 {\cal L}=\frac{x_t(y_{t+\delta t}-y_t)-y_t(x_{t+\delta t}-x_t)}{\delta t},
\end{equation}
which, upon substituting Eq.~\eqref{eq:discXY}, becomes
\begin{equation}\label{eq:Lconditional}
 {\cal L}=u(y_t^2-x_t^2)
 +\sqrt{\frac{2T_y}{\delta t}}x_t\eta_y
 -\sqrt{\frac{2T_x}{\delta t}}y_t\eta_x.
\end{equation}
At a fixed position, Eq.~\eqref{eq:Lconditional} is a linear combination of independent Gaussian variables plus a deterministic term. Thus its \emph{conditional} distribution is Gaussian, with mean $u(y^2-x^2)$ and variance $2 (T_yx^2+T_xy^2)/\delta t$. 
The corresponding conditional density, (denoted by the lowercase letter $p$ to distinguish it from the stationary PDF), is
\begin{equation}\label{eq:PLcond}
 p({\cal L}|x,y)=
 \frac{\sqrt{\delta t}}{2\sqrt{\pi(T_yx^2+T_xy^2)}}
 \exp\!\left[-\frac{\delta t\,[{\cal L}-u(y^2-x^2)]^2}
 {4(T_yx^2+T_xy^2)}\right].
\end{equation}
This density is defined for $(x,y)\ne(0,0)$; the single point at the origin does not affect the subsequent position integral. In the stationary state, the starting position is itself random. Averaging $p({\cal L}|x,y)$ over the stationary position PDF therefore gives
\begin{equation}\label{eq:PLstationary}
P_r^{\rm st}({\cal L}|X_0,Y_0)=\int_{-\infty}^{\infty} \int^{\infty}_{-\infty} dx\,dy\,
P_r^{\rm st}(x,y|X_0,Y_0)\, p({\cal L}|x,y).
\end{equation}
Here $(X_0,Y_0)$ specifies the \emph{fixed resetting point} in case A, whereas $(x,y)$ is the particle's position immediately before the observation step. In case B, we have to average the result over a circumference of a circle. Equation~\eqref{eq:PLstationary} is a mixture of Gaussians with position-dependent means and variances; the resulting stationary angular-momentum distribution is not generally Gaussian.

\subsection*{Mean angular momentum}
Averaging Eq.~\eqref{eq:Lconditional} first over the noises and then over the stationary position distribution gives
\begin{equation}\label{eq:Lmean2}
\overline{\cal L}=u\bigl(\overline{y^2}-\overline{x^2}\bigr).
\end{equation}
Note that the simple form in Eq. \eqref{eq:Lmean2} holds only for independent delta-correlated white  noises for which the cross-correlations
	with the opposite coordinate vanish in the It\^o convention: $\overline{x\eta_y}= \overline{y\eta_x}=0$. This is no longer the case for colored noises. In particular, for symmetric dichotomous noises the above cross-correlations do not vanish and provide an additional negative contribution to the mean angular momentum (see \cite{Herbeau2026}).

The same result follows from the local stationary probability current representation:
\begin{equation}\label{eq:LmeanCurrent}
\overline{\cal L}=\int_{-\infty}^{\infty} \int^{\infty}_{-\infty} dx\,dy\,
\bigl[xj_y^{(X_0,Y_0)}-yj_x^{(X_0,Y_0)}\bigr].
\end{equation}
Indeed, substituting Eq.~\eqref{j} into Eq.~\eqref{eq:LmeanCurrent}, the terms involving spatial derivatives are
$-T_y\int x\,\partial_yP_r^{\rm st}\,dx\,dy$
and $T_x\int y\,\partial_xP_r^{\rm st}\,dx\,dy$.
Integration by parts in $y$ and $x$, respectively, makes both terms vanish: the boundary terms vanish for a sufficiently rapidly decaying stationary PDF, while $\partial_y x=\partial_x y=0$. The remaining drift terms give Eq.~\eqref{eq:Lmean2}. Substitution of the second moments from Eq.~\eqref{eq:mom2} yields Eq.~\eqref{eq:Lmean} of the main text. 

\subsection*{Integral representation of the stationary PDF}
To evaluate Eq.~\eqref{eq:PLstationary}, we use the renewal representation \eqref{pr2}. For a given time $\tau$ since the last resetting event, its Gaussian propagator has the exponent appearing in Eq.~\eqref{pr2}. Introducing polar coordinates $x=\rho\cos\theta$, $y=\rho\sin\theta$, write that exponent as
$-Q(\tau,\theta)\rho^2-B(\tau,\theta)\rho-C(\tau)$, where
\begin{align}
	\begin{split}
		\label{eq:ABC}
Q(\tau,\theta)&=\frac{b\cos^2\theta+a\sin^2\theta+c\sin\theta\cos\theta}{4ab-c^2},\\
B(\tau,\theta)&=-\frac{2b\cos\theta\widetilde X_0+2a\sin\theta\widetilde Y_0+c(\cos\theta\widetilde Y_0+\sin\theta\widetilde X_0)}{4ab-c^2},\\
C(\tau)&=\frac{b\widetilde X_0^2+a\widetilde Y_0^2+c\widetilde X_0\widetilde Y_0}{4ab-c^2}.
\end{split}
\end{align}
Here $a,b,c$ and $\widetilde X_0,\widetilde Y_0$ are the functions of $\tau$ defined in Eqs.~\eqref{coeft} and \eqref{tilde}. 

The angular dependence of the noise variance is conveniently abbreviated as
\begin{equation}\label{eq:Sangle}
S(\theta)=T_y\cos^2\theta+T_x\sin^2\theta.
\end{equation}
Substituting the conditional density \eqref{eq:PLcond} into Eq.~\eqref{eq:PLstationary} gives
\begin{equation}\label{eq:PLpolar}
\begin{split}
P_r^{\rm st}({\cal L}|X_0,Y_0)
={}&\frac{r}{4\pi^{3/2}}\int_0^\infty
\frac{e^{-r\tau}\,d\tau}{\sqrt{4ab-c^2}}
\int_0^{2\pi}\frac{d\theta\,\sqrt{\delta t}}{\sqrt{S(\theta)}}
\int_0^\infty d\rho\,e^{-Q\rho^2-B\rho-C}\\
&\times\exp\!\left[-\frac{\delta t\,[{\cal L}+u\rho^2\cos(2\theta)]^2}
{4\rho^2S(\theta)}\right].
\end{split}
\end{equation}
Equation~\eqref{eq:PLpolar} is used to compute the angular-momentum PDFs numerically.

For circular resetting, the resetting position is $(R_0\cos\varphi_0,R_0\sin\varphi_0)$, with $\varphi_0$ uniform on $[0,2\pi)$. One first evaluates Eq.~\eqref{eq:PLpolar} at this resetting position and then averages over $\varphi_0$, as in Eq.~\eqref{eq:PLcircleaverage}. The same averaging of Eq.~\eqref{eq:Lmean2} yields the mean angular momentum presented in Eq.~\eqref{eq:Lcircle}. 

\subsection*{Derivation of the exponential tails}
We now derive the leading large-$|{\cal L}|$ decay rates, which are displayed in Eqs.~\eqref{eq:Ltails} and \eqref{eq:LcircleTail}. The crucial point to realise is that large angular momentum is associated with large distances from the origin, rather than with passages close to it. For fixed $\theta$, the noise factor in Eq.~\eqref{eq:PLpolar} contains $\exp[-\delta t\,{\cal L}^2/(4\rho^2S(\theta))]$ at small $\rho$; this suppresses small distances when $|{\cal L}|$ is large.

At large distances the leading quadratic part of the logarithm of the stationary position PDF is the same as for the BG without resetting. To see why, consider the renewal integral: at any fixed elapsed time $\tau$ the propagator is Gaussian, and as $\tau\to\infty$ its covariance approaches the stationary covariance of the BG without resetting. The corresponding quadratic cost is the smallest attainable one in the large-distance limit. The finite resetting position contributes terms at most linear in $\rho$ to the Gaussian exponent, while the resetting weight $e^{-r\tau}$ can be made subleading compared with $\rho^2$ by allowing $\tau$ to grow slowly with $\rho$. Hence, at leading logarithmic order,
\begin{equation}\label{eq:posGaussianTail}
\ln P_r^{\rm st}(\rho\cos\theta,\rho\sin\theta|X_0,Y_0)
=-q(\theta)\rho^2+o(\rho^2),
\qquad \rho\to\infty,
\end{equation}
where the positive quadratic coefficient is read directly from the stationary PDF without resetting, Eq.~\eqref{P} :
\begin{equation}\label{eq:qthetaL}
q(\theta)=A\left[
\left(T_y+\frac{u^2\Delta_T}{2}\right)\cos^2\theta+
\left(T_x-\frac{u^2\Delta_T}{2}\right)\sin^2\theta+
 u(T_x+T_y)\sin\theta\cos\theta\right].
\end{equation}

Set ${\cal L}=s |\cal L|$ with $|\cal L| \to \infty$, where  $s=+1$ or $-1$ distinguishes the two tails. The competition between the position cost $q(\theta)\rho^2$ and the noise cost $\delta t\,{\cal L}^2/[4\rho^2S(\theta)]$ implies $\rho^2=O({\cal L})$ . We therefore introduce $z=\rho^2/|{\cal L}| >0$. Retaining all terms proportional to $|\cal L|$ in the combined exponent gives
\begin{equation}\label{eq:FsL}
F_s(z,\theta)=q(\theta)z+
\frac{\delta t[\,s+uz\cos(2\theta)\,]^2}{4zS(\theta)}.
\end{equation}
Expanding the square separates the terms that depend on $z$:
\begin{equation}\label{eq:FsExpand}
F_s(z,\theta)=
\left[q(\theta)+\frac{\delta t\,u^2\cos^2(2\theta)}{4S(\theta)}\right]z
+\frac{\delta t}{4S(\theta)z}
+\frac{\delta t\,su\cos(2\theta)}{2S(\theta)}.
\end{equation}
Define $D(\theta)=q(\theta)+\delta t\,u^2\cos^2(2\theta)/[4S(\theta)]$, which is strictly positive for positive temperatures and a stable BG. The $z$-dependent part is $D(\theta)z+\delta t/[4S(\theta)z]$. Differentiating with respect to $z$ and setting the derivative to zero gives
\begin{equation}\label{eq:zstarL}
z_*(\theta)=\sqrt{\frac{\delta t}{4S(\theta)D(\theta)}}.
\end{equation}
At this minimum, the sum of the two $z$-dependent terms equals $\sqrt{\delta t\,D(\theta)/S(\theta)}$. The remaining integration over $\theta$ is governed by the smallest such exponent. We thus obtain the two decay rates
\begin{equation}\label{eq:lambdaLpm}
\lambda_s=\min_{0\le\theta<2\pi}\left[
\sqrt{\frac{\delta t}{S(\theta)}\left(q(\theta)+\frac{\delta t\,u^2\cos^2(2\theta)}{4S(\theta)}\right)}
+\frac{\delta t\,su\cos(2\theta)}{2S(\theta)}\right],
\qquad s=\pm1.
\end{equation}
Consequently, $\ln P({\cal L})=-\lambda_+{\cal L}+o({\cal L})$ as ${\cal L}\to+\infty$, and $\ln P({\cal L})=-\lambda_-|{\cal L}|+o(|{\cal L}|)$ as ${\cal L}\to-\infty$. This is the precise leading-logarithmic meaning of the exponential tails. The two rates can differ because the drift term in Eq.~\eqref{eq:FsExpand} changes sign with $s$.

Finally, the coefficient $q(\theta)$ is determined by the BG model without resetting, and neither $r$ nor the resetting position enters Eq.~\eqref{eq:lambdaLpm}. Uniform averaging over the resetting circle cannot change these leading exponential rates either. Resetting can, however, modify the amplitudes and subleading logarithmic corrections. This establishes the result in Eq.~\eqref{eq:LcircleTail} and explains why its leading rates do not depend on $R_0$ or $r$.

\section{Probability density function of the specific angular velocity}
\label{appendix:Pdew}

\subsection*{Conditional distribution and stationary PDF}
We use the same forward discretisation and the same notation as in Appendix~\ref{appendix:momentum}. The angular velocity is the angular momentum divided by the squared distance from the origin (i.e., the specific moment of inertia) at the beginning of the observation step:
\begin{equation}\label{eq:Wdis}
 {\cal W}=\frac{x_t(y_{t+\delta t}-y_t)-y_t(x_{t+\delta t}-x_t)}
 {\delta t(x_t^2+y_t^2)}.
\end{equation}
Substituting Eq.~\eqref{eq:discXY} gives
\begin{equation}\label{eq:Wconditional}
 {\cal W}=u\frac{y_t^2-x_t^2}{x_t^2+y_t^2}
 +\sqrt{\frac{2T_y}{\delta t}}\frac{x_t}{x_t^2+y_t^2}\eta_y
 -\sqrt{\frac{2T_x}{\delta t}}\frac{y_t}{x_t^2+y_t^2}\eta_x.
\end{equation}
Note that 
${\cal W}$ 
can be regarded as a superposition of
 two Gaussian noises $\eta_x$ and $\eta_y$. 
 Therefore, for every
 fixed $(x,y)\ne(0,0)$, ${\cal W}$ is Gaussian itself,
 with conditional mean $u (y^2-x^2)/(x^2+y^2)$ and variance $2 (T_yx^2+T_xy^2)/(\delta t (x^2+y^2)^2)$.
The stationary PDF is then the mixture
\begin{equation}\label{eq:PWaverage}
 P_r^{\rm st}({\cal W}|X_0,Y_0)
 =\int_{-\infty}^{\infty} \int_{-\infty}^{\infty} dx\,dy\,
 P_r^{\rm st}(x,y|X_0,Y_0)\,p({\cal W}|x,y) ,
\end{equation}
where
\begin{equation}\label{eq:PWcond}
	p({\cal W}|x,y)=
	\frac{\sqrt{\delta t}(x^2+y^2)}{2\sqrt{\pi(T_yx^2+T_xy^2)}}
	\exp\!\left[-\frac{\delta t(x^2+y^2)^2
		[\,{\cal W}-u(y^2-x^2)/(x^2+y^2)\,]^2}
	{4(T_yx^2+T_xy^2)}\right] \,.
\end{equation}

\subsection*{Integral representation}
We retain $S(\theta)$ from Eq.~\eqref{eq:Sangle} and the polar propagator coefficients $Q,B,C$ from Eq.~\eqref{eq:ABC}. In polar coordinates, Eq.~\eqref{eq:PWaverage} reads
\begin{equation}\label{eq:PWpolarSimple}
\begin{split}
P_r^{\rm st}({\cal W}|X_0,Y_0)
={}&\int_0^{2\pi}d\theta\int_0^\infty\rho\,d\rho\,
P_r^{\rm st}(\rho\cos\theta,\rho\sin\theta|X_0,Y_0)
\frac{\sqrt{\delta t}\,\rho}{2\sqrt{\pi S(\theta)}}\\
&\times\exp\!\left[-\frac{\delta t\rho^2
[\,{\cal W}+u\cos(2\theta)\,]^2}{4S(\theta)}\right].
\end{split}
\end{equation}
The two factors of $\rho$ in the radial integral arise, respectively, from the polar area element and the normalization of the conditional Gaussian. Inserting the renewal representation and combining the quadratic terms in the exponent, define
\begin{equation}\label{eq:A1W}
Q_{\cal W}(\tau,\theta)=Q(\tau,\theta)+
\frac{\delta t[\,{\cal W}+u\cos(2\theta)\,]^2}{4S(\theta)}.
\end{equation}
The radial integral is elementary:
\begin{equation}\label{eq:rho2integral}
\begin{split}
\int_0^\infty\rho^2e^{-Q_{\cal W}\rho^2-B\rho}\,d\rho
={}&\frac{\sqrt\pi(2Q_{\cal W}+B^2)e^{B^2/(4Q_{\cal W})}
\operatorname{erfc}\!\left(B/(2\sqrt{Q_{\cal W}})\right)-2\sqrt{Q_{\cal W}}B}
{8Q_{\cal W}^{5/2}}.
\end{split}
\end{equation}
We obtain the explicit two-fold integral
\begin{equation}\label{eq:PdeW}
\begin{split}
P_r^{\rm st}({\cal W}|X_0,Y_0)
={}&\frac{r}{4\pi^{3/2}}\int_0^\infty
\frac{e^{-r\tau-C}\,d\tau}{\sqrt{4ab-c^2}}
\int_0^{2\pi}d\theta\,
\sqrt{\frac{\delta t}{S(\theta)}}\\
&\times\frac{\sqrt\pi(2Q_{\cal W}+B^2)e^{B^2/(4Q_{\cal W})}
\operatorname{erfc}\!\left(B/(2\sqrt{Q_{\cal W}})\right)-2\sqrt{Q_{\cal W}}B}
{8Q_{\cal W}^{5/2}}.
\end{split}
\end{equation}
All coefficients inside the integrand depend on $\tau$ as specified above. This is the representation used for numerical evaluation of the angular-velocity PDFs.

\subsection*{Derivation of the inverse-cubic tail}
The origin of the angular-velocity tail is different from that of the angular-momentum tail. In Eq.~\eqref{eq:PWpolarSimple}, a large $|{\cal W}|$ suppresses contributions except from $\rho=O(|{\cal W}|^{-1})$. Suppose that the resetting positions do not include the origin, so that the stationary position density has a finite, nonzero limit $P_r^{\rm st}(0,0|X_0,Y_0)$ there. At leading order, this density can be replaced by its value at the origin, and the bounded deterministic term $u\cos(2\theta)$ can be neglected relative to ${\cal W}$. The remaining radial integral is
\begin{equation}\label{eq:WtailRadial}
\int_0^\infty\rho^2\exp\!\left[-\frac{\delta t\,{\cal W}^2}{4S(\theta)}\rho^2\right]d\rho
=\frac{2\sqrt\pi\,S(\theta)^{3/2}}{\delta t^{3/2}|{\cal W}|^3}.
\end{equation}
Combining this with the normalization in Eq.~\eqref{eq:PWpolarSimple} gives
\begin{equation}\label{eq:WtailIntermediate}
P_r^{\rm st}({\cal W}|X_0,Y_0)
\sim\frac{P_r^{\rm st}(0,0|X_0,Y_0)}{\delta t\,|{\cal W}|^3}
\int_0^{2\pi}S(\theta)\,d\theta.
\end{equation}
Since the angular integral is $\pi(T_x+T_y)$, this yields Eq.~\eqref{eq:Wtail} of the main text. The same calculation applies to circular resetting, with the stationary position density at the origin replaced by its circular-resetting counterpart, which gives Eq.~\eqref{eq:WcircleTail}.

\subsection*{Mean angular velocity and circular resetting}
Averaging Eq.~\eqref{eq:Wconditional} over the noises at fixed $(x,y)$, followed by averaging over the stationary position PDF, gives the mean angular velocity in Eq.~\eqref{eq:angular4}. The mean exists because the singularity near the origin is integrable in two dimensions, even though the second moment does not exist.

For numerical evaluation, one may insert the renewal representation and perform the radial integral. With the coefficients $Q,B,C$ defined in Eq.~\eqref{eq:ABC}, this gives
\begin{equation}\label{eq:WmeanExplicit}
\begin{split}
\overline{\cal W}={}&-\frac{ur}{2\pi}
\int_0^\infty\frac{d\tau\,e^{-r\tau-C}}{\sqrt{4ab-c^2}}
\int_0^{2\pi}d\theta\,\cos(2\theta)\\
&\times\left[\frac{1}{2Q}-\frac{\sqrt\pi B}{4Q^{3/2}}
 e^{B^2/(4Q)}\operatorname{erfc}\!\left(\frac{B}{2\sqrt Q}\right)\right].
\end{split}
\end{equation}
The bracket is the integral $\int_0^\infty\rho\,e^{-Q\rho^2-B\rho}\,d\rho$. Equation~\eqref{eq:WmeanExplicit} therefore provides a two-fold integral representation without first computing the full angular-velocity PDF.
For case B, the resetting position is chosen uniformly on the circle at every resetting event. The stationary PDF and the mean angular velocity are obtained by averaging the corresponding fixed-point expressions over the resetting position, as in Eqs.~\eqref{eq:WcircleMean} and \eqref{eq:WcircleTail}.

\section{A series representation of Eqs.  \eqref{eq:margcircleReal}.}
\label{C}

Here we concentrate on the inner integrals in Eqs. \eqref{eq:margcircleReal} seeking a convenient series representation. Write the function in the exponential in the first line in  Eqs. \eqref{eq:margcircleReal} as
\begin{equation}
	-\frac{(x-\ell \cos\varphi)^2}{4a}
	=-\frac{x^2}{4a}-\frac{\ell^2}{8a}
	+\frac{x\ell}{2a}\cos\varphi-\frac{\ell^2}{8a}\cos(2\varphi).
\end{equation}
Introduce next the generalised modified Bessel
function of two arguments \cite{Dattoli1990}
\begin{equation}\label{eq:genBesselDef}
	{\cal I}_0(A,B)
	\equiv \frac{1}{2\pi}\int_0^{2\pi}
	\exp[A\cos\varphi+B\cos(2\varphi)]\,d\varphi .
\end{equation}
This function has the exact
representation in terms of modified Bessel functions \cite{Dattoli1990}
\begin{equation}\label{eq:genBesselSeries}
	{\cal I}_0(A,B)
	=\sum_{n=-\infty}^{\infty}I_{2n}(A)I_n(B)
	=I_0(A)I_0(B)+2\sum_{n=1}^{\infty}I_{2n}(A)I_n(B).
\end{equation}
Thus the angular integration in Eq.~\eqref{eq:margcircleReal} can be performed
completely, and the marginal PDFs reduce to the one-dimensional renewal
integrals
\begin{align}
	\begin{split}\label{eq:pxcircleExact}
		P_{r}^{\rm st}(x|R_0)
		&=r\int_0^\infty d\tau\,e^{-r\tau}
		\frac{\exp[-x^2/(4a)-\ell^2/(8a)]}{\sqrt{4\pi a}}\,
		{\cal I}_0\!\left(\frac{x \ell}{2a},-\frac{\ell^2}{8a}\right) , \\
		P_{r}^{\rm st}(y|R_0)
		&=r\int_0^\infty d\tau\,e^{-r\tau}
		\frac{\exp[-y^2/(4b)-\ell^2/(8b)]}{\sqrt{4\pi b}}\,
		{\cal I}_0\!\left(\frac{y\ell}{2b},-\frac{\ell^2}{8b}\right) .
	\end{split}
\end{align}
\end{document}